\documentclass[conference,letterpaperu]{IEEEtran}

\usepackage[utf8]{inputenc} 
\usepackage[T1]{fontenc}
\usepackage{url}
\usepackage{ifthen}
\usepackage{cite}
\usepackage[cmex10]{amsmath} 
\usepackage{url}
\usepackage{amsfonts}
\usepackage{array}
\usepackage{xargs}   
\usepackage[pdftex,dvipsnames]{xcolor}
\usepackage{mathtools}
\usepackage[font=footnotesize]{caption}
\usepackage{relsize}
\usepackage{hyperref}
\usepackage{subfig}
\usepackage{acro}
\usepackage{bm}

\usepackage{tikz,pgfplots}
\pgfplotsset{compat=1.18} 
\usepgfplotslibrary{groupplots}

\definecolor{dgreen}{RGB}{1,50,32}
\definecolor{lightGreen}{rgb}{0, 1, .5}
\definecolor{lightPink}{rgb}{1, .71, .76}
\definecolor{cornflowerBlue}{RGB}{100, 149, 237}
\definecolor{navajoWhite}{rgb}{1, 0.87, 0.68}
\definecolor{midnightBlue}{rgb}{0.1, 0.1, 0.44}
\definecolor{purple}{rgb}{.63, .13, .94}
\definecolor{darkSeaGreen}{rgb}{.56, 0.74, 0.56}

\DeclarePairedDelimiter\abs{\lvert}{\rvert}%
\newcommand{\vv}{\ensuremath{\,\vert\,}}
\newcommand{\cc}{\ensuremath{,\,}}
\newcommand{\expnumber}[2]{{#1}\mathrm{e}{#2}}

\newcommand{\pdel}{\ensuremath{p_D}}
\newcommand{\pins}{\ensuremath{p_I}}
\newcommand{\psub}{\ensuremath{p_S}}
\newcommand{\ptrans}{\ensuremath{p_T}}
\newcommand{\sbin}{\ensuremath{\{ 0, 1\}}}

\newcommand{\ygen}{\ensuremath{\bm{x}}}
\newcommand{\ygenIx}{\ensuremath{x}}

\newcommand{\yin}{\ensuremath{\bm{x}^{\text{in}}}}
\newcommand{\yout}{\ensuremath{\bm{x}^{\text{out}}}}

\newcommand{\yinIx}{\ensuremath{x^{\text{in}}}}

\newcommand{\youtIx}{\ensuremath{x^{\text{out}}}}

\newcommand{\Yval}{\ensuremath{\xi}}

\newcommand{\YStateVal}{\ensuremath{\bm{\zeta}}}
\newcommand{\YStateValIx}{\ensuremath{\zeta}}
\newcommand{\mess}{\ensuremath{\bm{m}}}

\newcommand{\ngen}{\ensuremath{n}}
\newcommand{\nin}{\ensuremath{n_{\text{in}}}}

\newcommand{\nout}{\ensuremath{n_{\text{out}}}}

\newcommand{\predyout}{\ensuremath{\bm{\hat{x}}^{\text{out}}}}
\newcommand{\predyin}{\ensuremath{\bm{\hat{x}}^{\text{in}}}}

\newcommand{\rec}{\ensuremath{\bm{r}}}

\newcommand{\recIx}{\ensuremath{r}}

\newcommand{\recbpsk}{\ensuremath{\bm{r}_{\text{bpsk}}}}

\newcommand{\nrecbpsk}{\ensuremath{\nout}} 

\newcommand{\nrec}{\ensuremath{n_{\text{rec}}}}

\newcommand{\dfrom}{\ensuremath{d}}
\newcommand{\dto}{\ensuremath{d^{\prime}}}

\newcommand{\bipolar}{\ensuremath{\phi}}
\newcommand{\syndrome}{\ensuremath{\text{syn}}}
\newcommand{\bin}{\ensuremath{\text{bin}}}

\newcommand{\ybipolar}{\ensuremath{\bm{x}^{\bipolar}}}

\newcommand{\ymodel}{\ensuremath{\bm{\hat{x}}}}
\newcommand{\ymodelIx}{\ensuremath{\hat{x}}}

\newcommand{\marker}{\ensuremath{\bm{s_{\text{m}}}}}
\newcommand{\markerFreq}{\ensuremath{N_{\text{m}}}}

\newcommand{\gen}{\ensuremath{\bm{G}}}
\newcommand{\pc}{\ensuremath{\bm{H}}}
\newcommand{\attQuery}{\ensuremath{\bm{Q}}}
\newcommand{\attKey}{\ensuremath{\bm{K}}}
\newcommand{\attValue}{\ensuremath{\bm{V}}}

\newcommand{\alphabetSize}{\ensuremath{q}}

\newcommand{\bcjrformerInput}{\ensuremath{\bm{Y}}}
\newcommand{\bcjrformerInputIx}{\ensuremath{Y}}

\newcommand{\bcjrformerInputBit}{\ensuremath{\bm{Y^{\text{symb}}}}}

\newcommand{\bcjrformerInputState}{\ensuremath{\bm{Y^{\text{state}}}}}
\newcommand{\bcjrformerInputStateIx}{\ensuremath{Y^{\text{state}}}}

\newcommand{\normalizedInner}{\ensuremath{\tilde{\bm{x}}^\text{out}}}

\newcommand{\convn}{\ensuremath{n_\text{c}}}

\newcommand{\convk}{\ensuremath{k_\text{c}}}

\newcommand{\alphabet}{\ensuremath{\mathbb{F}_q}}

\newcommand{\noutput}{\ensuremath{N_{\YStateValIx}}}

\newcommand{\offset}{\ensuremath{\bm{o}}}
\newcommand{\offsetIx}{\ensuremath{o}}

\newcommand{\Ndec}{\ensuremath{N_{\text{dec}}}}

\newcommand{\hiddenDim}{\ensuremath{d_{\text{k}}}}

\newcommand{\nheads}{\ensuremath{n_{\text{h}}}}

\acsetup{single=true, single-style=long}

\DeclareAcronym{ecct}{
    short = ECCT,
    long = Error Correcting Code Transformer,
}
\DeclareAcronym{ldpc}{
    short = LDPC,
    long = low-density parity-check,
}

\DeclareAcronym{map}{
    short = MAP,
    long = maximum a posteriori,
    first-style=long,
    subsequent-style=long
}
\DeclareAcronym{bcjr}{
    short = BCJR,
    long = Bahl-Cocke-Jelinek-Raviv,
}
\DeclareAcronym{bp}{
    short = BP,
    long = Belief Propagation,
    first-style=long,
    subsequent-style=long,
}
\DeclareAcronym{dna}{
    short = DNA,
    long = Deoxyribonucleic acid,
    first-style= short,
    subsequent-style=short,
}

\DeclareAcronym{rnn}{
    short = RNN,
    long = recurrent neural network
}
\DeclareAcronym{bigru}{
    short = BiGRU,
    long = bidirectional gated recurrent unit
}

\DeclareAcronym{ids}{
    short = IDS,
    long = {insertion, deletion, and substitution}
}

\DeclareAcronym{bpsk}{
    short = BPSK,
    long = binary phase-shift keying,
    first-style=long,
    subsequent-style=long
}

\DeclareAcronym{biso}{
    short = BISO,
    long = {binary-input symmetric-output},
    first-style=long,
    subsequent-style=long
}

\DeclareAcronym{bce}{
    short = BCE,
    long = binary cross-entropy,
    first-style=long,
    subsequent-style=long
}

\DeclareAcronym{ber}{
    short = BER,
    long = bit error rate,
    subsequent-style=long
}
\DeclareAcronym{ser}{
    short = SER,
    long = symbol error rate,
    subsequent-style=long
}

\usetikzlibrary {decorations.pathreplacing, arrows, calc,positioning,shapes.misc, graphs, matrix, quotes,backgrounds, fit, intersections, }

\tikzstyle{terminal} = [draw=black, line width=.5pt, rectangle, rounded corners = 2pt, minimum height = 10pt, fontscale=-1, top color=white, bottom color=gray!20]
   
\tikzstyle{container} = [draw=black!40, rectangle, very thick, inner sep=0.1cm,  rounded corners=.5mm, top color=white, bottom color=gray!30]

\tikzstyle{vector} = [draw=black, rectangle, rounded corners=2pt, minimum width=50pt, minimum height=10pt]
\tikzset{
fontscale/.style = {font=\relsize{#1}}
}

\tikzstyle{receivedNode} = [draw=black, line width=.5pt, rectangle, rounded corners = 2pt, minimum height = 10pt]
   
\tikzstyle{embeddingNode} = [
 line width=.5pt, minimum height = 14pt, minimum width=20pt, inner sep=1pt, append after command={
   \pgfextra
        \draw[sharp corners]
    (\tikzlastnode.west)%
    [rounded corners=0pt] |- (\tikzlastnode.north)%
    [rounded corners=0pt] -| (\tikzlastnode.east)%
    [rounded corners=2pt] |- (\tikzlastnode.south)%
    [rounded corners=0pt] -| (\tikzlastnode.west);
   \endpgfextra
   }
]
\tikzstyle{vnode}=[circle, draw, very thick, minimum size=5pt, top color=white, bottom color=gray!20]
\tikzstyle{cnode}=[rectangle,minimum size=6mm,rounded corners=1mm,
                          very thick,draw=black, top color=gray!20, bottom color=gray!60]

\tikzstyle{posEmbeddingNode} = [line width=.5pt, minimum height = 3pt, minimum width=23.5pt, inner sep=0pt, append after command={
   \pgfextra
        \draw[sharp corners, fill=darkSeaGreen!80]%
    (\tikzlastnode.west)%
    [rounded corners=2pt] |- (\tikzlastnode.north)%
    [rounded corners=2pt] -| (\tikzlastnode.south east)%
    [rounded corners=0pt] |- (\tikzlastnode.south)%
    [rounded corners=0pt] -| (\tikzlastnode.west);
   \endpgfextra
   }]

\tikzstyle{seqEmbeddingNode} = [line width=.5pt, rectangle, rounded corners = 1pt, minimum height = 14pt, minimum width=3pt, inner sep=0pt, append after command={
   \pgfextra
        \draw[sharp corners, fill=cornflowerBlue!80 ]
    (\tikzlastnode.west)%
    [rounded corners=0pt] |- (\tikzlastnode.north)%
    [rounded corners=0pt] -| (\tikzlastnode.east)%
    [rounded corners=0pt] |- (\tikzlastnode.south)%
    [rounded corners=2pt] -| (\tikzlastnode.west);
   \endpgfextra
   }
]
\tikzstyle{layerNode} = [draw=black, line width=.5pt, rectangle, rounded corners = 2pt, minimum height = 10pt]

\tikzstyle{symbolReprNode} = [draw=black, rectangle, minimum height=12pt, minimum width=4pt, inner sep=0]

\tikzstyle{stateReprNode} = [draw=black, rectangle, minimum height=12pt, minimum width=8pt, inner sep=0]

\tikzstyle{outputReprNode} = [draw=black, rectangle, minimum height=4pt, minimum width=4pt, inner sep=0]

\tikzstyle{addNode} = [
    circle,
    draw=black,
    inner sep=0,
    minimum size=7pt,
    path picture={
      \draw [black]
            (path picture bounding box.90) -- (path picture bounding box.270)
            (path picture bounding box.0) -- (path picture bounding box.180);
    }
]

\pgfplotsset{every axis legend/.append style={anchor=north,font=\footnotesize}}

\begin{document}
\title{Transformer-Based Decoding in Concatenated Coding Schemes Under Synchronization Errors} 

\author{%
\IEEEauthorblockN{Julian Streit, Franziska Weindel and Reinhard Heckel}%
\IEEEauthorblockA{\\
School of Computation, Information and Technology, Technical University of Munich \\
Munich Center for Machine Learning \\
Email: julian.streit@tum.de, franziska.weindel@tum.de, reinhard.heckel@tum.de} 
}%


\maketitle
\thispagestyle{plain}
\pagestyle{plain}
\begin{abstract}
We consider the reconstruction of a codeword from multiple noisy copies that are independently corrupted by insertions, deletions, and substitutions.
This problem arises, for example, in \ac{dna} data storage.
A common code construction uses a concatenated coding scheme that combines an outer linear block code with an inner code, which can be either a nonlinear marker code or a convolutional code.
Outer decoding is done with \ac{bp}, and inner decoding is done with the \ac{bcjr} algorithm.
However, the \ac{bcjr} algorithm scales exponentially with the number of noisy copies, which makes it infeasible to reconstruct a codeword from more than about four copies.
In this work, we introduce \textit{BCJRFormer}, a transformer-based neural inner decoder. \textit{BCJRFormer} achieves error rates comparable to the \ac{bcjr} algorithm for binary and quaternary single-message transmissions of marker codes. Importantly, \textit{BCJRFormer} scales quadratically with the number of noisy copies. This property makes \textit{BCJRFormer} well-suited for \ac{dna} data storage, where multiple reads of the same \ac{dna} strand occur. To lower error rates, we replace the \ac{bp} outer decoder with a transformer-based decoder. Together, these modifications yield an efficient and performant end-to-end transformer-based pipeline for decoding multiple noisy copies affected by insertion, deletion, and substitution errors. Additionally, we propose a novel cross-attending transformer architecture called \textit{ConvBCJRFormer}. This architecture extends \textit{BCJRFormer} to decode transmissions of convolutional codewords, serving as an initial step toward joint inner and outer decoding for more general linear code classes. 
\end{abstract}
\acresetall
\section{Introduction}
\ac{dna} data storage is an emerging storage medium that encodes binary data into \ac{dna} sequences for high-density, long-term storage. However, writing, storing, and reading \ac{dna} are error-prone processes that produce multiple noisy reads of the same sequence~\cite{heckelCharacterizationDNAData2019}. Errors can be categorized into two categories: (1) substitutions and erasures, and (2) insertions and deletions. While optimal linear codes are known for erasures and substitutions, no similar guarantees exist for insertions and deletions.

A common approach to address insertion, deletion, and substitution errors is to use a concatenated coding scheme that uses an inner and an outer encoder/decoder pair. The inner code is often nonlinear\,---\,using, for example, marker codes~\cite{sellersBitLossGain1962} or watermark codes~\cite{daveyReliableCommunicationChannels2001}\,---\,and is typically decoded with the \ac{bcjr} algorithm. Under the assumption of perfect channel state information, the \ac{bcjr} algorithm achieves \ac{map} decoding. 
The outer code is a linear block code, such as a \ac{ldpc} code~\cite{wangSymbolLevelSynchronizationLDPC2011a} 
or a polar code~\cite{talPolarCodesDeletion2022}, 
and is decoded using \ac{bp}. \ac{bp} achieves \ac{map} decoding when the code's Tanner graph has no cycles~\cite{tannerRecursiveApproachLow1981}.
\begin{figure}[t!]
    \centering
    \begin{tikzpicture}[scale=1]        
    \node (r1)[receivedNode, minimum width=22pt] at (0,0) {$\rec^1$};

    \node (channel) [receivedNode, fill=gray!40, left = .5cm of r1] {\footnotesize IDS Channel};
    \node (channelToReceived) [coordinate, left=1pt of r1.west] {};
    \path (channel.east) edge[-latex] (channelToReceived);
    \node (codeword) [left = .3cm of channel.west] {$\yin$};
    \path (codeword.east) edge[-latex] (channel.west);


    \node (r2)[receivedNode, minimum width=33pt, right = 4pt of r1.east] {$\rec^2$};
    \node (r3)[receivedNode, minimum width=25pt, right = 4pt of r2.east] {$\rec^3$};
    \node (r4)[receivedNode, minimum width=33pt, right = 4pt of r3.east] {$\rec^4$};
    \node (r5)[receivedNode, minimum width=22pt, right = 4pt of r4.east] {$\rec^5$};

    \node(window) [receivedNode, minimum width=156pt, above= .4cm of r3.north, fill=navajoWhite] {Drift Sliding Window};

    \node(linearEmbedding) [above = (.5cm) of channel, align=center, fontscale=1, xshift=-5pt] {\footnotesize Linear \\[-1ex] \footnotesize Embedding};

    \node(y1SeqEmbedding)[seqEmbeddingNode, above=.4cm of window.north west, xshift=6pt] {};
    \node (y1)[embeddingNode, right = 0pt of y1SeqEmbedding.east] {$\bcjrformerInput^1$};
    \node (y1PosEmbedding)[posEmbeddingNode, above=0pt of y1.north, xshift=-1.75pt] {};

    \node(Embeddings) [above=0pt of linearEmbedding,align=center, fontscale=1] {\footnotesize \textcolor{darkSeaGreen!80}{\textbf{Positional}} \& \textcolor{cornflowerBlue!80}{\textbf{Sequence}}\\[-1ex]\footnotesize Embedding};

    \node(y2SeqEmbedding)[seqEmbeddingNode, right=7pt of y1.east] {};
    \node (y2)[embeddingNode, right = 0pt of y2SeqEmbedding.east] {$\bcjrformerInput^2$};
    \node (y2PosEmbedding)[posEmbeddingNode, above=0pt of y2.north, xshift=-1.75pt] {};

    \node(y3SeqEmbedding)[seqEmbeddingNode, right=7pt of y2.east] {};
    \node (y3)[embeddingNode, minimum width=20pt, right=0pt of y3SeqEmbedding.east] {$\bcjrformerInput^3$};
    \node (y3PosEmbedding)[posEmbeddingNode, above=0pt of y3.north, xshift=-1.75pt] {};

    \node(y4SeqEmbedding)[seqEmbeddingNode, right=7pt of y3.east] {};
    \node (y4)[embeddingNode, right=0pt of y4SeqEmbedding.east] {$\bcjrformerInput^4$};
    \node (y4PosEmbedding)[posEmbeddingNode, above=0pt of y4.north, xshift=-1.75pt] {};

    \node(y5SeqEmbedding)[seqEmbeddingNode, right=7pt of y4.east] {};
    \node (y5)[embeddingNode, right=0pt of y5SeqEmbedding.east] {$\bcjrformerInput^5$};
    \node (y5PosEmbedding)[posEmbeddingNode, above=0pt of y5.north, xshift=-1.75pt] {};

    \node(transformer) [receivedNode, minimum width=156pt, above= .4cm of y3PosEmbedding.north, fill=lightPink] {Transformer};



    \node (meanAgg) [draw, circle, fontscale=-2, inner sep=0.5pt, fill=gray!10] at (channel |- transformer) {$\frac{1}{M}\!\!\sum$};

    \node (output) at (codeword |- meanAgg) {$\predyin$};

    \node (r1window)[coordinate, above=.4cm of r1.north] {};
    \path (r1) edge[-latex] (r1window);

    \node (r2window)[coordinate, above=.4cm of r2.north] {};
    \path (r2) edge[-latex] (r2window);

    \node (r3window)[coordinate, above=.4cm of r3.north] {};
    \path (r3) edge[-latex] (r3window);

    \node (r4window)[coordinate, above=.4cm of r4.north] {};
    \path (r4) edge[-latex] (r4window); 
    
    \node (r5window)[coordinate, above=.4cm of r5.north] {};
    \path (r5) edge[-latex] (r5window);

    \node (y1FromWindow)[coordinate, below=14.5pt of y1PosEmbedding.south] {};
    \node (windowy1)[coordinate, below=(11.5pt) of y1FromWindow] {};
    \path (windowy1) edge[-latex] node[circle, draw, minimum size=1pt, inner sep=0, fill, yshift=-1pt] (circ1) {} (y1FromWindow);

    \node (y2FromWindow)[coordinate, below=14.5pt of y2PosEmbedding.south] {};
    \node (windowy2)[coordinate, below=11.5pt of y2FromWindow] {};
    \path (windowy2) edge[-latex] node[circle, draw, minimum size=1pt, inner sep=0, fill, yshift=-1pt] {} (y2FromWindow);
    
    \node (y3FromWindow)[coordinate, below=14.5pt of y3PosEmbedding.south] {};
    \node (windowy3)[coordinate, below=11.5pt of y3FromWindow] {};
    \path (windowy3) edge[-latex] node[circle, draw, minimum size=1pt, inner sep=0, fill, yshift=-1pt] {} (y3FromWindow);
    
    \node (y4FromWindow)[coordinate, below=14.5pt of y4PosEmbedding.south] {};
    \node (windowy4)[coordinate, below=11.5pt of y4FromWindow] {};
    \path (windowy4) edge[-latex] node[circle, draw, minimum size=1pt, inner sep=0, fill, yshift=-1pt] {} (y4FromWindow);
    
    \node (y5FromWindow)[coordinate, below=14.5pt of y5PosEmbedding.south] {};
    \node (windowy5)[coordinate, below=11.5pt of y5FromWindow] {};
    \path (windowy5) edge[-latex] node[circle, draw, minimum size=1pt, inner sep=0, fill, yshift=-1pt] {} (y5FromWindow);

    \node(linearEmbeddingDashedLineFrom) [coordinate] at ([xshift=-10pt]circ1.west) {}; 
    \node(linearEmbeddingDashedLine) [coordinate, right=145pt of linearEmbeddingDashedLineFrom] {};
    \path (linearEmbeddingDashedLineFrom) edge[dashed] (linearEmbeddingDashedLine);


    \node(linEmbToDashedLine) [coordinate] at ([xshift=-7pt]linearEmbeddingDashedLineFrom.west) {};
    \path (linearEmbedding) edge[-latex] (linEmbToDashedLine);


    \node(SeqEmbFromEmbeddings) [coordinate] at ([xshift=-7pt]y1SeqEmbedding.west) {};

    \node(EmbeddingsToSeqEmb) [coordinate] at ([yshift=7pt]linearEmbedding.north east) {};

    \path (EmbeddingsToSeqEmb) edge[-latex] (SeqEmbFromEmbeddings);
    
    \node (y1transformer)[coordinate, above=(.4cm) of y1PosEmbedding.north] {};
    \path (y1PosEmbedding) edge[-latex] (y1transformer);
    
    \node (y2transformer)[coordinate, above=(.4cm) of y2PosEmbedding.north] {};
    \path (y2PosEmbedding) edge[-latex] (y2transformer);
    
    \node (y3transformer)[coordinate, above=(.4cm) of y3PosEmbedding.north] {};
    \path (y3PosEmbedding) edge[-latex] (y3transformer);
    
    \node (y4transformer)[coordinate, above=(.4cm) of y4PosEmbedding.north] {};
    \path (y4PosEmbedding) edge[-latex] (y4transformer);
    
    \node (y5transformer)[coordinate, above=(.4cm) of y5PosEmbedding.north] {};
    \path (y5PosEmbedding) edge[-latex] (y5transformer);

    \path (transformer) edge [-latex] (meanAgg);

    \path (meanAgg) edge[-latex] (output);

    \end{tikzpicture}
    \caption{Overview of \textit{BCJRFormer} for jointly decoding marker sequences over the IDS channel. }
    \label{fig:BCJRFormerOverview}
\end{figure}
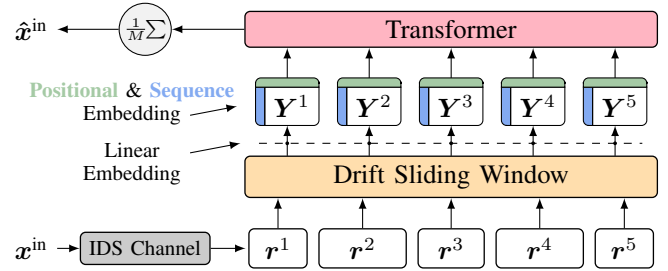

The computational complexity of the \ac{bcjr} algorithm scales exponentially with the number of input sequences, which makes joint decoding over multiple noisy copies infeasible. To address this limitation, we propose \textit{BCJRFormer}, a transformer-based architecture for inner decoding of marker codes. The architecture achieves error rates close to those of joint decoding with the \ac{bcjr} algorithm while scaling only quadratically with the number of noisy copies. 

To further reduce bit error rates, we replace the outer \ac{bp} algorithm with the \ac{ecct} proposed in~\cite{choukrounErrorCorrectionCode2022b}.
The authors show that their \ac{ecct} decoder outperforms both \ac{bp} and other neural decoders on \ac{biso} channels. By adapting the input to \ac{ecct}, we propose a two-step decoding approach for \ac{ldpc} codes concatenated with a marker code. This approach reduces \aclp{ber} and improves the efficiency of decoding for correcting insertion, deletion, and substitution errors compared to traditional \ac{bp} and \ac{bcjr} decoding.

Separating inner and outer decoding is necessary because incorporating the state-space of general linear codes into the \ac{bcjr} algorithm is computationally infeasible. Convolutional codes are an exception because they exhibit a sparse diagonal structure~\cite{bahlOptimalDecodingLinear1974}. As an initial step toward joint inner and outer decoding of more general linear code classes, we extend \textit{BCJRFormer} to convolutional codes by using a cross-attention mechanism that incorporates their state structure. We demonstrate that this decoder, named \textit{ConvBCJRFormer}, achieves error rates only slightly worse than \ac{bcjr} decoding while offering a generalizable mechanism to incorporate linear code information. 

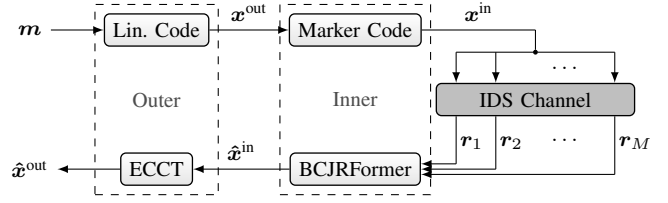
\begin{figure}[t]
    \centering
    \begin{tikzpicture}[scale=1]        

        \pgfdeclarelayer{background}
        \pgfsetlayers{background,main} 
    
        \node[fontscale=-1] (m)  at (0, 0) { $\mess$} ; 
        \node (linear) [terminal,right = 20pt of m.east, fontscale=-1] {Lin. Code};
        \node (marker) [terminal, right= 30pt of linear.east ] {Marker Code};

        \node (markerToChannel) [coordinate, right=43pt of marker] {};
        \node (channel) [receivedNode, below = 20pt of markerToChannel, minimum width=75pt, fill=gray!50, fontscale=-1] {IDS Channel};
        
        \node (bcjrformer) [terminal, below = 40pt of marker] {BCJRFormer};
        \node (ecct) [terminal] at (bcjrformer -| linear) {ECCT};
    
        \node[fontscale=-1] (dec) at (ecct -| m) {$\predyout$};
    
        \path (m) edge[-latex] (linear);
    
        \path (linear) edge[-latex] node[above, fontscale=-1] { $\yout$ } (marker);
    
        \path(bcjrformer) edge[-latex]node[above, fontscale=-1] { $\predyin$ } (ecct);
    
        \path(ecct) edge [-latex]node[above, fontscale=-1] {} (dec);

        \path (marker) edge[-]node[auto, fontscale=-1] { $\yin$ }(markerToChannel);

        \node (edgeSep) [draw, circle, fill=black, inner sep=.5pt, below=7.5pt of markerToChannel] {};

        \path(markerToChannel) edge[-] (edgeSep);

        \node (m1ToCh) [coordinate, left=30pt of edgeSep.center] {};
        \node (m2ToCh) [coordinate, left=15pt of edgeSep.center] {};
        \node (m3ToCh) [coordinate, right=11pt of edgeSep.center] {};
        \node (m4ToCh) [coordinate, right=30pt of edgeSep.center] {};

        \path (edgeSep) edge[-] (m1ToCh);
        \path (edgeSep) edge[-] (m4ToCh);

        \path (m1ToCh) edge[-latex] ([xshift=-30pt]channel.north);        \path(m2ToCh) edge[-latex] ([xshift=-15pt]channel.north);
        
        \path(m3ToCh) -- node[midway, fontscale=-1]{$\hdots$} ([xshift=11pt]channel.north);
        \path(m4ToCh) edge[-latex] ([xshift=30pt]channel.north);

        \node (r1FromCh) [coordinate, left=30pt of channel.south] {};
        \node (r2FromCh) [coordinate, left=15pt of channel.south] {};
        \node (r3FromCh) [coordinate, right=11pt of channel.south] {};
        \node (r4FromCh) [coordinate, right=30pt of channel.south] {};

        \node (BCJRFromr1) [coordinate, above=2pt of bcjrformer.east] {};
        \node (BCJRFromr2) [coordinate, above= 0pt of bcjrformer.east] {};      \node (BCJRFromr3) [coordinate, above= 0pt of bcjrformer.east] {};
        \node (BCJRFromr4) [coordinate, below=2pt of bcjrformer.east] {};

        \node (BCJRForm1Con) [coordinate] at (r1FromCh |- BCJRFromr1) {};
        \path (r1FromCh) edge[-] node[right, fontscale=-1, xshift=-2pt, yshift=-1pt]{$\rec_1$}  (BCJRForm1Con);
        \path (BCJRForm1Con) edge[-latex] (BCJRFromr1);
         
        \node (BCJRForm2Con) [coordinate] at (r2FromCh |- BCJRFromr2) {};
        \path (r2FromCh) edge[-] node[right, fontscale=-1, xshift=-2pt]{$\rec_2$}  (BCJRForm2Con);
        \path (BCJRForm2Con) edge[-latex] (BCJRFromr2);

        \node (BCJRForm3Con) [coordinate] at (r3FromCh |- BCJRFromr3) {};
        \path (r3FromCh) -- node[midway, fontscale=-1, yshift=1pt]{$\hdots$} (BCJRForm3Con);
        
        \node (BCJRForm4Con) [coordinate] at (r4FromCh |- BCJRFromr4) {};
        \path (r4FromCh) edge[-] node[right, fontscale=-1, xshift=-2pt, yshift=1pt]{$\rec_M$} (BCJRForm4Con);
        \path (BCJRForm4Con) edge[-latex] (BCJRFromr4);

        \node (containerInner) [fit=(bcjrformer) (marker), draw, dashed, label={[anchor=center, text=black!70]center:\footnotesize{Inner}}] {};
        \node (containerOuter) [fit=(ecct) (linear), draw, dashed, label={[anchor=center, text=black!70]center:\footnotesize{Outer}}] {};
    \end{tikzpicture}
    \caption{Joint decoding of concatenated codes using \textit{BCJRFormer} as the inner decoder and ECCT as the outer decoder.}
    \label{fig:FullApproach}
\end{figure}

We summarize our contributions as follows:\footnote{Our code is available at \url{https://github.com/streit-julian/BCJRFormer}.}
\begin{itemize}
    \item We introduce a sliding window input representation derived from the \ac{bcjr} algorithm. This approach yields \textit{BCJRFormer}, a transformer model that achieves error rates comparable to those of the \ac{bcjr} algorithm in single-sequence transmissions.
    \item We extend \textit{BCJRFormer} to jointly decode multiple copies of transmitted sequences and demonstrate that the resulting decoder achieves error rates similar to the \ac{bcjr} algorithm while operating with only quadratic complexity.
    \item We demonstrate that ECCT can serve as an outer decoder in concatenated coding schemes, achieving lower error rates than \ac{bp}.
    \item By combining \textit{BCJRFormer} as the inner decoder with ECCT as the outer decoder (see Figure~\ref{fig:FullApproach}), we propose a transformer-based pipeline for end-to-end decoding of concatenated codewords over the \ac{ids} channel that outperforms approaches based on \ac{bcjr} or \ac{bp} decoders.
    \item We introduce \emph{ConvBCJRFormer}, a cross-attending transformer architecture that extends \textit{BCJRFormer} to decode convolutional codes as an initial step toward transformer-based joint inner and outer decoding of linear codeword transmissions over the \ac{ids} channel.
\end{itemize}



\section{Related Work}
In \ac{dna} data storage, multiple noisy reads of the same sequence are often clustered and then aligned to yield a candidate sequence for error correction~\cite{organickRandomAccessLargescale2018, barlev2024deepdnastoragescalable, girsch2025trace,antkowiakLowCostDNA2020}. Because the \ac{bcjr} algorithm becomes infeasible for larger cluster sizes due to its exponential complexity, the article~\cite{maaroufConcatenatedCodesMultiple2023a} analyzes decoding convolutional codes by applying the \ac{bcjr} algorithm to each read independently and then multiplying the resulting a posteriori probabilities. The paper~\cite{srinivasavaradhanTrellisBMACoded2021} builds on this approach by proposing Trellis BMA\,\textemdash\,an algorithm that combines separate trellis calculations with bitwise majority alignment. 

The article~\cite{barlev2024deepdnastoragescalable} proposes DNAformer, an end-to-end retrieval solution for \ac{dna} data storage that also uses a transformer-based architecture to reconstruct data from multiple noisy reads. Their work focuses on the challenge of imperfect clustering, which can result in reads coming from different sequences. 
A recent paper~\cite{girsch2025trace} formulates the problem of reconstructing an uncoded \ac{dna} sequence from multiple noisy copies as a next token prediction task and evaluates decoder-only transformers for this purpose. Although the tasks differ, we note that their transformers have millions of parameters, whereas our \textit{BCJRFormer} has fewer than one million.

Several neural decoders have been proposed to replace iterative decoding algorithms\,\textemdash\,such as \ac{bcjr} or \ac{bp}\,\textemdash\,across various channel models~\cite{nachmaniLearningDecodeLinear2016a, nachmaniDeepLearningMethods2018, sazliNeuralNetworkImplementation2007,shlezingerDatadrivenFactorGraphs2020b, farsadDatadrivenSymbolDetection2021, shlezingerViterbiNetSymbolDetection2019}. We distinguish between model-based and model-free architectures~\cite{shlezingerModelbasedDeepLearning2023}.

In the paper~\cite{maDeepLearningBasedDetection2024b}, the authors propose two model-based \ac{rnn} architectures\,\textemdash\,FBNet and FBGRU\,\textemdash\,to decode insertion, deletion, and substitution errors in marker codes. Their models outperform the \ac{bcjr} algorithm when the channel state information is imperfect. The input for their models is derived from the \ac{bcjr} algorithm, much like how we design the input for our \textit{BCJRFormer}.

The authors of the article~\cite{kargiDeepLearningBased2024b} propose a \ac{bigru} decoder to address vanishing and exploding gradient issues in \acp{rnn}. They use one \ac{bigru} network for inner decoding of marker codes and a second \ac{bigru} network for outer decoding of codewords encoded with an \ac{ldpc} code or a convolutional code to correct deletion and substitution errors. 

Model-free networks are characterized by more general neural architectures, but are often harder to design because the network must learn both the code structure and the reconstruction process. The authors of the paper~\cite{choukrounErrorCorrectionCode2022b} propose using the attention mechanism in transformer networks to embed linear code information directly into the model architecture. They introduce \ac{ecct}, which restricts the attention mechanism to attend only to those input bits which are related by parity-check equations. \ac{ecct} outperforms existing neural decoders on memoryless \ac{biso} channels. 

More recently, the article~\cite{park2025crossmpt} proposes CrossMPT to improve the decoding performance of \ac{ecct}. CrossMPT is also transformer-based and uses two masked cross-attention modules to handle channel noise and syndrome, imitating the check and variable node updates of the \ac{bp} algorithm. We employ a similar cross-attention mechanism to decode inner convolutional codes; however, we derive the mask from the code's generator matrix rather than from the parity-check matrix, which is used in CrossMPT's approach. 

\section{Background}
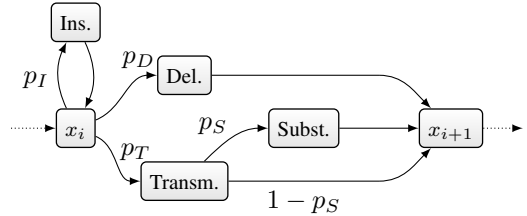
\begin{figure}[t]
    \centering
    \begin{tikzpicture}[
                        node distance=5mm,
                        text height=1.5ex,text depth=.25ex]
      \matrix[row sep=5pt,column sep=15pt] {
        &\node (ins) [terminal] {Ins.}; & & &  && \\
         && \node (del) [terminal] {Del.}; & & \node (delToyi1) [coordinate] {}; && \\
         \node(toyi) [coordinate] {}; &\node (yi) [terminal] {$\ygenIx_i$}; & & \node (subs) [terminal] {Subst.}; &  &  \node (yi1) [terminal] {$\ygenIx_{i+1}$};& \node(fromyi1)[coordinate] {};\\
         && \node (trans) [terminal] {Transm.};& & \node(transToyi1) [coordinate] {}; && \\    
      };
    
        \graph [edge quotes=auto] {
            (toyi) ->[densely dotted, >=latex] (yi);
            (yi) ->["$\pins$", out=115, in=-115, >=latex] (ins) ->[out=-65, in=65, >=latex] (yi);
            (yi) ->["$\pdel$", out=22.5, in=180, edge quotes={above=2pt, near end}, >=latex] (del) -- (delToyi1) ->[out=0, in=135, >=latex] (yi1);
            (yi) ->["$\ptrans$", out=-22.5, in=180, edge quotes={right=1pt, near start}, >=latex] (trans);
            (trans) -- ["$1 - \psub$", edge quotes=below, >=latex] (transToyi1) ->[out=0, in=-135, >=latex] (yi1);
            (trans) ->["$\psub$", out=45, in=180, edge quotes={above left=-2pt}, >=latex] (subs) ->[, >=latex] (yi1);
            (yi1) ->[densely dotted, >=latex] (fromyi1);
        };
    \end{tikzpicture}
    \caption{State transitions for a codeword symbol $\ygenIx_i$ transmitted through the \acs{ids} channel.}
    \label{fig:ids-channel}
\end{figure}
This section presents an overview of the key concepts relevant to our work. We first introduce our channel model and describe concatenated coding schemes, with a focus on the \ac{bcjr} algorithm that is used to derive the input for both \textit{BCJRFormer} and \textit{ConvBCJRFormer}. We then discuss the transformer architecture and the original formulation of the \ac{ecct} decoder for \ac{biso} channels.
\subsection{Channel Model}
We consider transmission via the \acl{ids} channel over a finite field $\alphabet$ of order $q = 2^p$, where $p$ is a positive integer. 
The \ac{ids} channel is represented as a state machine, as we show in Figure~\ref{fig:ids-channel}. Each symbol $\ygenIx_i$ in a sequence $\ygen \in \alphabet^\ngen$ enters the state machine independently. The symbol is deleted with probability $\pdel$. With insertion probability $\pins$, the channel outputs a random symbol and resets its state. With transmission probability $\ptrans = 1 - \pins - \pdel$, the symbol is transmitted. In this case, the symbol is substituted with probability $\psub$ by a different symbol chosen uniformly at random~\cite{daveyReliableCommunicationChannels2001}.

\subsection{Concatenated Schemes and Linear Codes}
\label{sec:linear_code}
Concatenated coding schemes consist of an inner encoder/decoder pair and an outer encoder/decoder pair. A message $\mess \in \alphabet^{k}$ is first transformed by the outer encoder into an outer codeword $\yout \in \alphabet^{\nout}$. This outer codeword is then encoded by the inner encoder to produce the inner codeword $\yin \in \alphabet^{\nin}$. After transmission, the inner decoder synchronizes the received sequence $\rec \in \alphabet^{\nrec}$ and computes log-likelihood values that serve as soft information for the outer decoder. The outer decoder then uses this soft information, together with the structure of the outer code, to correct any remaining substitution errors.

The outer code is typically a linear block code. A $(k, \nout)$ linear block code is defined by its generator matrix $\gen \in \alphabet^{k \times \nout}$. Multiplying the message $\mess$ by the generator matrix produces the codeword $\yout = \mess\gen$. The parity-check matrix \( \pc \in \alphabet^{(\nout-k) \times \nout} \) is defined so that $\pc \yout = \bm{0}$ holds for every codeword. The parity-check matrix can be represented by a Tanner graph, a bipartite graph in which the rows of the parity-check matrix serve as check nodes and the columns as variable nodes~\cite{tannerRecursiveApproachLow1981}. A check node and a variable node are connected by a weighted edge if the corresponding entry of the parity-check matrix is non-zero, with the edge weight given by that entry. 

The \ac{bp} algorithm operates on the Tanner graph by iteratively passing soft information between check and variable nodes to reconstruct the original message. \ac{bp} yields maximum a posteriori predictions of the codeword if the Tanner graph is cycle-free. If cycles are present, \ac{bp} still provides effective approximations. In this work, we focus on \ac{ldpc} outer codes~\cite{gallagerLowdensityParitycheckCodes1962a}. \ac{ldpc} codes are a class of linear codes designed for efficient decoding via message passing.
\subsection{Convolutional Codes}
\label{sec:bgConvCode}
\begin{figure}[t]
\centering
\begin{tikzpicture}[scale=.8]
  \begin{axis}[
    axis line style = {ultra thick, black},
    axis equal image,
    enlargelimits=false,
    xtick=\empty,
    ytick=\empty,
    colormap={binary}{color(0cm)=(white); color(1cm)=(black!90!cornflowerBlue!95)},
    point meta min=0, 
    point meta max=1,
  ]
    \addplot [
      matrix plot,
      mesh/rows=9,
      mesh/cols=18, 
      point meta=explicit,
      draw=black,
    ]
    table [x index=1,y index=0, meta index=2] 
    {cross_mask.txt};
  \end{axis}
\end{tikzpicture}
    \caption{Generator matrix $G \in \sbin^{9 \times 18}$ for the $(2, 1, 2)$ convolutional code with generator polynomials $[5, 7]_8$, used for an outer codeword with $\nout = 7$. White entries indicate unmasked positions.}
    \label{fig:generator_conv}
\end{figure}
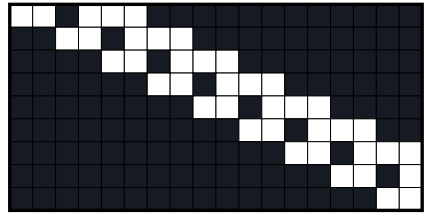
A $(\convn, \convk, m)$ convolutional code with rate $\convk / \convn$ and memory $m$ is described by $\convk$ generator polynomials $g = [g_1, g_2, \ldots g_{\convk}],$ where each $g_l \in \alphabet[x]^{\convn}$ is a vector of $\convn$ polynomials with degree at most $m$. For brevity, we express the polynomials in octal notation. For example, we denote a pair of generator polynomials as \([5,7]_8\), where each digit represents the binary coefficients of a polynomial in octal form. For example, \([5,7]_8\) represents the polynomials \(101\) and \(111\). 

Each generator polynomial $g_l$ is applied to one of the $\convk$ input symbols. At time $i$, the encoder uses a sliding window of $m +1$ inputs\,---\,the current input and the $m$ preceding inputs \,---\, to produce $\convn$ output symbols. This process can be viewed as a Markov chain: the encoder's state is given by the contents of its memory, so transitions from time $i$ to time $i+1$ depend only on the current state and the new input.

For fixed transmission lengths, the decoder benefits from knowing the final memory state. To provide this information, we zero-terminate the code by appending the zero vector $\bm{0}^m$ to the codeword. Consequently, given an outer codeword with length $\nout$, the inner codeword has length $\nin = \frac{\convn}{\convk}(\nout + m)$. 

An inner convolutional code can also be interpreted as a linear block code. In this perspective, the generator polynomials define a structured generator matrix $G$ that maps an input vector (augmented by the $m$ zero-termination symbols) to the codeword. The structure of $G$ is typically banded, with each row being a shifted version of the generator polynomials. Figure~\ref{fig:generator_conv} shows the generator matrix for a rate $1/2$ convolutional code with generator polynomials $[5, 7]_8$ as an example. 

After encoding, we add a pseudorandom offset $\offset \in \mathbb{F}_2^{\nin}$ to the convolutional codeword. This offset, which is known to the decoder, improves error rates by mitigating the cyclic structure of convolutional codes~\cite{buttigiegImprovedBitError2015}.
 
\subsection{Inner Code Construction and BCJR} 
 \label{sec:bgMarkerCodesBCJR}

Inner marker codes insert a fixed marker sequence \( \marker \in \alphabet^{n_{\text{m}}} \) at a fixed interval \( \markerFreq \) between the symbols of the outer codeword. The resulting inner codeword has length \( \nin = \nout + n_{\text{m}} \lfloor\frac{\nout}{\markerFreq}\rfloor\). These markers enable the inner decoder to synchronize the received sequence $\rec$.

For each inner codeword symbol $\yinIx_i$ (where $1 \leq i \leq \nin$), the inner decoder estimates the a posteriori probabilities
\begin{equation}
\label{eq:bgBCJRAPosterioriProbabilities}
P(\yinIx_i = \Yval \vv \rec)  \propto P(\yinIx_i = \Yval, \rec),
\end{equation}
where $\Yval \in \alphabet = \{0, \ldots, q-1\}$. 
The \ac{bcjr} algorithm is a forward-backward method that leverages the \ac{ids} channel's Markov property to compute these probabilities. Assuming that the symbols of the outer codeword are independent and identically distributed, the \ac{bcjr} algorithm yields exact results. In this work, we derive the algorithm specifically for marker codes.

To elicit the \ac{ids} channel's Markov property, we follow the paper~\cite{daveyReliableCommunicationChannels2001} and introduce a latent drift variable \( D_i \), defined as the difference between the number of insertions and deletions that occur before transmitting the \( i \)-th symbol. Transmitting the $i$-th symbol results in a state transition from state $D_i$ to $D_{i + 1}$. Each state transition emits between $0$ and $I_{\max} + 1$ symbols, where $I_{\max}$ is a parameter that limits the number of consecutive insertions per symbol to reduce computational complexity.  Consequently, if $D_i = \dfrom$, then $D_{i+1}$ can take any value in the set $\{ \dfrom- 1\cc \dfrom\cc \dfrom + 1\cc \ldots\cc \dfrom + I_{\max}\}$. Using the Markov property, we rewrite the joint probabilities as 
\begin{equation*}
 P(\yinIx_i = \Yval \cc \rec) = \sum_{\dfrom}\sum_{\dto = \dfrom - 1}^{\dfrom + I_{\max}} P(\yinIx_i = \Yval \cc \rec, \dfrom, \dto),  
\end{equation*}
where the first summation is over all possible realizations $\dfrom$ of the drift variable $D_{i-1}$. We factorize the term in the second summation as follows:
\begin{align}
\begin{split}
\label{eq:bjcr_markov}
    P(\yinIx_i = \Yval \cc \rec, \dfrom, \dto) = &P(\rec_1^{i + \dfrom - 1}, \dfrom)\\ &\cdot P(\yinIx_i = \Yval \cc \rec_{i + \dfrom}^{i + \dto}, \dto \vv \dfrom) \\ &\cdot 
    P(\rec_{i + \dto + 1}^{\nin} \vv \dto),
\end{split}
\end{align}
where we denote by $\rec_a^b$ the sequence $\recIx_a \cc \recIx_{a+1}\cc \ldots \cc \recIx_b$ of the vector $\rec$. We abbreviate the three factors in Equation~\eqref{eq:bjcr_markov} in order of appearance as $\alpha_{i-1}(\dfrom)$, $\gamma_i(\dfrom\cc \dto)$, and $\beta_i(\dto)$.
We deduce \ac{bcjr}'s forward (\( \alpha_i \)) and backward (\( \beta_i \)) recursions as
\begin{align*}
\begin{split}
\alpha_{i}(\dto) = &\frac{\pins}{q}\alpha_i(\dto-1) + \pdel \alpha_{i-1}(\dto+1) \\ 
&+ \ptrans \alpha_{i-1}(\dto) \sum_{\Yval \in \alphabet}P(\yinIx_i = \Yval)F(\Yval, \recIx_{i + \dto}),
\end{split}
\end{align*}
and
\begin{align*}
\begin{split}
\beta_{i}(\dfrom) = &\frac{\pins}{q}\beta_i(\dfrom +1) + \pdel \beta_{i+1}(\dfrom+1) \\ 
&+ \ptrans \beta_{i+1}(\dfrom) \sum_{\Yval \in \alphabet}P(\yinIx_i = \Yval)F(\Yval, \recIx_{i + 1 + \dfrom}),
\end{split}
\end{align*}
where
$$
F(\Yval, \recIx_i) = 
\begin{cases}
    \frac{\psub}{q-1} & \text{if } \Yval \neq \recIx_{i} ,\\ 
    1 - \psub & \text{else.}
\end{cases}
$$
Since the initial drift ($0$) and final drift ($\nrec - \nin$) are known, the initial conditions for both recursions are 
\begin{align*}
    \alpha_0(\dto) = \begin{cases}
        1  & \text{if } \dto = 0, \\ 
        0 & \text{else,}
    \end{cases} 
\end{align*}
and 
\begin{align*}
    \beta_{\nin}(\dfrom)  = \begin{cases}
        1 & \text{if }\dfrom = \nrec - \nin, \\
        0 & \text{else.}
    \end{cases}
\end{align*}
By combining the forward and backward recursions with the branching metrics $\gamma_i$, we simplify the calculation of the joint probabilities in Equation~\eqref{eq:bgBCJRAPosterioriProbabilities} to obtain
\begin{align*}
    P(\yinIx_i = \Yval \cc \rec) = \sum_{\dfrom} &\frac{\pins}{q} \alpha_i(\dfrom-1)\beta_i(\dfrom) \\[-.7em]
    &+ \pdel \alpha_{i-1}(\dfrom + 1) \beta_i(\dfrom) \\ 
    &+ \ptrans \alpha_{i-1}(\dfrom)\beta_i(\dfrom) F(\Yval, \recIx_{i+\dfrom}).
\end{align*}
For a more detailed discussion of the \ac{bcjr} algorithm, we refer the reader to the works~\cite{bahlDecodingChannelsInsertions1975a, jelinekStatisticalMethodsSpeech1998}. The \ac{bcjr} algorithm can also decode convolutional inner codes over the \ac{ids} channel by considering all combinations of convolutional and drift states~\cite{maaroufConcatenatedCodesMultiple2023a}.

\subsection{Transformer and ECCT}
\label{sec:TransformerAndECCT}
The transformer architecture uses attention to capture dependencies across the input sequence~\cite{vaswaniAttentionAllYou2017a}. Let \( \hiddenDim \) denote the model's embedding dimension. An attention layer is defined as
\begin{equation}
\label{eq:attention}
Att(\attQuery, \attKey, \attValue) = \text{Softmax}\left(\frac{\attQuery \attKey^T}{\sqrt{\hiddenDim}}\right)\attValue,
\end{equation}
where the query $\attQuery$, key $\attKey$, and value $\attValue$ matrices are learned from the input tokens. If all three matrices are derived from the same input sequence, this is called self-attention. Each attention layer consists of $\nheads$ attention heads. Each head performs attention on a reduced dimension of $\hiddenDim/\nheads$. The outputs of all heads are concatenated and then passed through a final linear layer.

A transformer stacks multiple sequential layers. Each layer comprises an attention block followed by a feed-forward block. In the attention block, the transformer first normalizes its input, then applies multi-head attention, and finally adds a residual connection. After processing all layers, the transformer projects the final output to a task-specific dimension.

The \ac{ecct} was proposed for decoding transmissions over a \ac{biso} channel, a type of memory-free channels that introduces substitution errors independently of the binary input symbol~\cite{choukrounErrorCorrectionCode2022b}. 

\ac{ecct} incorporates linear code information by masking unrelated bit positions in its self-attention module. The model's input is constructed to be independent of the transmitted codeword. 

Before transmission over a \ac{biso} channel, we modulate the codeword using \ac{bpsk}. In this modulation, each bit is first mapped to a bipolar value\,---\,specifically, $0$ is mapped to $1$ and $1$ is mapped to $-1$.
Let $\recbpsk \in [-1, 1]^{\nrecbpsk}$ denote the received sequence corresponding to this \ac{bpsk}-modulated codeword. The model's input is formed by concatenating the magnitude vector $\abs{\recbpsk} \in \mathbb{R}^{\nrecbpsk}$ with the syndrome vector $\bipolar(\syndrome(\bin(\recbpsk))) \in \{-1, 1\}^{\nrecbpsk - k}$. Here, we define the syndrome of a sequence as $\syndrome(\bm{x}) = \pc \bm{x} \in \mathbb{F}_2^{\nrecbpsk - k}$, the bi\-polar mapping as $\bipolar(\bm{x}) = 1 - 2\bm{x}$, and the binarization as $\bin(\bm{x}) = 0.5(1 - \text{sign}(\bm{x}))$.

By concatenating these two vectors, the model captures the channel's multiplicative noise independently of the transmitted codeword, which enables training using only the all-zero sequence~\cite{bennatanDeepLearningDecoding2018b}.

\section{Method}

In this section, we introduce \textit{BCJRFormer}, a neural decoder for efficiently decoding multiple noisy copies of inner codewords encoded with marker codes. We also describe our modifications to the \ac{ecct} model for use as an outer decoder alongside \textit{BCJRFormer}, which leads to lower error rates for transmissions over the \ac{ids} channel. We then propose the \textit{ConvBCJRFormer} architecture, an extension of \textit{BCJRFormer} to jointly synchronize and decode convolutional codes.

\subsection{BCJRFormer}
\label{sec:BCJRFormer}
Our proposed \textit{BCJRFormer} model, illustrated in Figure~\ref{fig:BCJRFormerOverview}, is a decoder-only transformer. We focus on constructing an appropriate input representation that achieves near-\ac{bcjr} performance and scales to jointly decode multiple noisy copies of the same codeword. We separately describe the input construction for the single copy scenario and the multiple copy scenario.

\subsubsection{Single Copy}
\label{sec:methBCJRFormerSingleTransmission}

For $1 \leq i \leq \nin$, we define input tokens $\bcjrformerInput_i \in \mathbb{R}^{\delta \times q}$, over an alphabet of size $q$ and a sliding window with size $\delta = d_{\max} - d_{\min} + 1$, where $d_{\min}$ and $d_{\max}$ serve as the lower and upper bounds on the drift of all received sequences. For each drift value $j$ satisfying $d_{\min} \leq j \leq d_{\max}$ and for each alphabet symbol $\Yval \in \alphabet$, we define 
\begin{equation}
     \label{eq:BCJRFormerInput}
    \bcjrformerInputIx_{i,j,\Yval} = P(\yinIx_i = \Yval)F(\Yval, \recIx_{i + d_j}).
 \end{equation} 
Assuming that the outer codeword symbols are independent and identically distributed, the prior probability for any non-marker symbol \( \yinIx_i \) is \( P(\yinIx_i = \Yval) = \frac{1}{q} \). Consequently, the input tokens are $\bcjrformerInputIx_{i, j, \Yval} = \frac{1}{q}$ for all alphabet values $\Yval$. 

Let $i_{\text{m}}$ denote the position of an inserted marker symbol \( \yinIx_{i_{\text{m}}} \). Because the decoder knows the value \( \Yval_{\text{m}} \) of this marker symbol, we get the prior probability \( P(\yinIx_{i_{\text{m}}} = \Yval_{\text{m}}) = 1 \). Then the input tokens resolve to
$$
 \bcjrformerInputIx_{i_{\text{m}},j, \Yval} = \begin{cases}
     F(\Yval_{\text{m}}, \recIx_{i_{\text{m}} + d_j}) &\text{if } \Yval = \Yval_{\text{m}}, \\
     0 & \text{else.}
 \end{cases}
$$

We flatten each token \( \bcjrformerInput_{i} \) and then embed the flattened tokens into the transformer's hidden dimension \( \hiddenDim \) using a linear layer. Because the transformer is permutation-invariant, we add learned positional embeddings to each embedded token before passing them through the transformer. The transformer's output, with dimensions \( \nin \times \hiddenDim \), is passed through a final linear layer of size \( \hiddenDim \times 1 \) and then through the sigmoid function \( \sigma(x) = \frac{1}{1 + \exp(-x)} \) to produce the prediction probabilities \( \predyin \in [0, 1]^{\nin}\).

\subsubsection{Multiple Sequence Alignment}
\label{sec:bcjrformer-msa}
The input representation extends naturally to handle multiple noisy copies of a transmitted codeword. Let \( M \) denote the number of transmissions. For the $k$-th transmission (with \( 1 \leq k \leq M \)) and for symbol position $1\leq i \leq \nin$, we define the token matrices \( \bcjrformerInput_i^k \in \mathbb{R}^{\delta \times q}\) as in Equation~\eqref{eq:BCJRFormerInput} and apply the same embedding. We then concatenate the tensors $\bcjrformerInput^k$ along their first dimension and add a second positional embedding that encodes the sequence position $k$. In Appendix~\ref{sec:appSequenceEmbedding}, we demonstrate that this sequence position encoding reduces error rates. 

The embedded input passes through the transformer and a final \( \hiddenDim \times 1 \) linear layer, as in the single-sequence case, yielding tokens $\bm{\hat{y}} \in \mathbb{R}^{M\nin}$. We compute \( \bm{\hat{y}}^{\text{in}} \in \mathbb{R}^{\nin} \) as
\begin{equation}
    \label{eq:mean_dim_reduction}
    \hat{y}^{\text{in}}_i = \frac{1}{M} \sum_{j=1}^M \hat{y}_{j\nin + i},
\end{equation}
since we found that simple averaging performs as well as a linear layer (see Appendix~\ref{sec:appMultipleSequenceAggregation}). 
Finally, \( \bm{\hat{y}}^{\text{in}} \) is passed through the sigmoid function $\sigma$ to yield prediction probabilities $\predyin$.

\subsection{ECCT as Outer Decoder}
\label{sec:MethodologyOuterECCT}
We adapt \ac{ecct} for use as an outer decoder by transforming its input. We normalize the inner decoder's approximations of the a posterior probabilities $P(\yinIx_i = 0 \vv \rec)$ (as introduced in Equation~\eqref{eq:bgBCJRAPosterioriProbabilities}), which yields a probabilistic vector $\normalizedInner \in [0, 1]^n$. We then transform this vector into bipolar vectors defined by $\ybipolar = \bipolar\left(\normalizedInner\right) \in [-1, 1]^{\nout}$. Then the input to \ac{ecct} is the concatenation of the magnitude part $\abs{\ybipolar}$ and the syndrome part $\bipolar\left(\syndrome\left(\bin\left(\ybipolar\right)\right)\right)$.

The inner decoder introduces noise that depends on the specific input codeword. As a result, outer decoding over the IDS channel is influenced by the codeword, unlike decoding over a \ac{biso} channel where the noise is the same regardless of the input. For example, aligning an all-zero codeword is simpler than aligning a codeword with alternating patterns. Consequently, we cannot train only on the zero codeword as done in the article~\cite{choukrounErrorCorrectionCode2022b}; instead, we train on pseudorandomly generated codewords to capture the codeword-dependent noise effects.

\subsection{ConvBCJRFormer}
\begin{figure}
    \centering
\begin{tikzpicture}[font=\footnotesize, scale=.1]       


\node(rec) at (0,0) {$\bm{r}$};

\node[coordinate, above = 6pt of rec.north] (coordRecSplit){};
\node[coordinate, left = 35pt of coordRecSplit] (coordRecSplitLeft){};
\node[coordinate, right = 35pt of coordRecSplit] (coordRecSplitRight){};

\matrix(symbolInput)[row sep=0pt,column sep=2pt, matrix of nodes, nodes={symbolReprNode}, nodes in empty cells, above = 10pt of coordRecSplitLeft] {
 & & & & |[draw=none]| $\ldots$& & & & \\
};
\matrix(stateInput)[row sep=0pt,column sep=2pt, matrix of nodes, nodes={stateReprNode}, nodes in empty cells, above = 10pt of coordRecSplitRight] {
 & &|[draw=none]| $\ldots$ & &  \\
};

\node (symbolLinEmb)[layerNode, fill=darkSeaGreen!50, above = 15pt of symbolInput] {Embedding};
\node (stateLinEmb)[layerNode, fill=darkSeaGreen!50, above = 15pt of stateInput] {Embedding};

\node (symbolPosEmbAdd)[addNode, above = 25pt of symbolLinEmb] {};
\node (statePosEmbAdd)[addNode, above = 25pt of stateLinEmb] {};

\node(symbolPosEmb)[layerNode, fill=gray!30, left = 10pt of symbolPosEmbAdd] {Pos. Embedding};
\node(statePosEmb)[layerNode, fill=gray!30,right = 10pt of statePosEmbAdd] {Pos. Embedding};


\node (symbolSelfAttention)[layerNode, fill=lightPink, above = 17pt of symbolPosEmbAdd, align=center] {Self-Attention \\ Transformer};
\node (stateSelfAttention)[layerNode, fill=lightPink, above = 17pt of statePosEmbAdd, align=center] {Self-Attention \\ Transformer};

\node (symbolResidualStart)[coordinate, above=4pt of symbolSelfAttention] {};
\node (symbolResidualP1)[coordinate, left = 40pt of symbolResidualStart] {};

\node (symbolLayerNorm)[layerNode, above = 8pt of symbolResidualStart] {Norm};
\node (symbolLayerCrossAttn)[layerNode, fill=navajoWhite, above = 11pt of symbolLayerNorm, align = center] {Masked \\ Cross-Attention};

\node (symbolResidualAdd) [addNode, above =10pt of symbolLayerCrossAttn] {};
\node (symbolResidualP2)[coordinate, left=40pt of symbolResidualAdd.center] {};
\node (symbolGeneratorMask)[layerNode, dashed, fill=purple!30, right = 10pt of symbolResidualAdd, anchor=west] {$\gen^T$};
\node(symbolIntersectionGeneratorCrossAttn)[coordinate] at (symbolLayerCrossAttn.north -| symbolGeneratorMask) {};

\node(symbolIntersectionCrossAttnState)[coordinate] at (symbolLayerCrossAttn.east -| stateSelfAttention) {};
\node(symbolKeyValueSplit)[coordinate, left = 15pt of symbolIntersectionCrossAttnState] {};

\node(symbolKeyPathStart)[coordinate, above = 5pt of symbolKeyValueSplit] {};
\node(symbolKeyPathEnd)[coordinate] at (symbolLayerCrossAttn.east |- symbolKeyPathStart) {};

\node(symbolValuePathStart)[coordinate, below = 5pt of symbolKeyValueSplit] {};
\node(symbolValuePathEnd)[coordinate] at (symbolLayerCrossAttn.east |- symbolValuePathStart) {};

\node (symbolResidual2Start)[coordinate, above=4pt of symbolResidualAdd] {};
\node (symbolResidual2P1)[coordinate, left = 40pt of symbolResidual2Start] {};
\node (symbolLayerNorm2)[layerNode, above = 8pt of symbolResidual2Start] {Norm};
\node (symbolFeedForwardLayer)[layerNode, fill= cornflowerBlue!40, above= 7pt of symbolLayerNorm2] {FFNN};

\node (symbolResidual2Add) [addNode, above =7pt of symbolFeedForwardLayer] {};
\node (symbolResidual2P2)[coordinate, left=40pt of symbolResidual2Add.center] {};

\node (stateResidualStart)[coordinate, above=50pt of symbolIntersectionCrossAttnState] {};
\node (stateResidualP1)[coordinate, right = 40pt of stateResidualStart] {};

\node (stateLayerNorm)[layerNode, above = 8pt of stateResidualStart] {Norm};
\node (stateLayerCrossAttn)[layerNode, fill=navajoWhite, above = 11pt of stateLayerNorm, align = center] {Masked \\ Cross-Attention};

\node (stateResidualAdd) [addNode, above =10pt of stateLayerCrossAttn] {};
\node (stateResidualP2)[coordinate, right=40pt of stateResidualAdd.center] {};
\node (stateGeneratorMask)[layerNode, dashed, fill=purple!30, left = 10pt of stateResidualAdd, anchor=east] {$\gen$};
\node(stateIntersectionGeneratorCrossAttn)[coordinate] at (stateLayerCrossAttn.north -| stateGeneratorMask) {};

\node(stateIntersectionCrossAttnSymbol)[coordinate] at (stateLayerCrossAttn.west -| symbolResidual2Add) {};
\node(stateKeyValueSplit)[coordinate, right = 15pt of stateIntersectionCrossAttnSymbol] {};

\node(stateKeyPathStart)[coordinate, above = 5pt of stateKeyValueSplit] {};
\node(stateKeyPathEnd)[coordinate] at (stateLayerCrossAttn.west |- stateKeyPathStart) {};

\node(stateValuePathStart)[coordinate, below = 5pt of stateKeyValueSplit] {};
\node(stateValuePathEnd)[coordinate] at (stateLayerCrossAttn.west |- stateValuePathStart) {};

\node (stateResidual2Start)[coordinate, above=4pt of stateResidualAdd] {};
\node (stateResidual2P1)[coordinate, right = 40pt of stateResidual2Start] {};
\node (stateLayerNorm2)[layerNode, above = 8pt of stateResidual2Start] {Norm};
\node (stateFeedForwardLayer)[layerNode, fill= cornflowerBlue!40, above= 7pt of stateLayerNorm2] {FFNN};

\node (stateResidual2Add) [addNode, above =7pt of stateFeedForwardLayer] {};
\node (stateResidual2P2)[coordinate, right=40pt of stateResidual2Add.center] {};


\node (symbolOut)[coordinate] at (stateIntersectionCrossAttnSymbol |- stateResidual2Add.north) {};

\node (symbolLayerNormHead)[layerNode, above=10pt of symbolOut] {Norm};
\node (symbolFeedForwardHead)[layerNode, fill=cornflowerBlue!80, above=7pt of symbolLayerNormHead] {FC};

\node (stateLayerNormHead)[layerNode, above=10pt of stateResidual2Add.north] {Norm};
\node (stateFeedForwardHead)[layerNode, fill=cornflowerBlue!80, above=7pt of stateLayerNormHead] {FC};

\node[coordinate] (fullOutputAlignment) at ($(stateFeedForwardHead)!0.5!(symbolFeedForwardHead)$) {};

\matrix(fullOutput)[row sep=0pt,column sep=1pt, matrix of nodes, nodes={outputReprNode, fill=black!60}, nodes in empty cells, above = 17pt of fullOutputAlignment] {
 & & & & & & & & & & & & & & &  & & & |[fill=gray!60]|  & |[fill=gray!60]|  & |[fill=gray!60]| & |[fill=gray!60]| & |[fill=gray!60]| & |[fill=gray!60]| & |[fill=gray!60]| & |[fill=gray!10]| & |[fill=gray!10]| \\
};

\node (symbolFullOutputIntersection)[coordinate] at (symbolLayerNormHead |- fullOutput.south) {};
\node (stateFullOutputIntersection)[coordinate] at (stateLayerNormHead |- fullOutput.south) {};

\matrix(lossIllustration)[row sep=0pt,column sep=.5pt, matrix of nodes, nodes={outputReprNode}, nodes in empty cells, above = 10pt of fullOutput] {
 |[minimum width=96pt, fill=black!60](yIn)| & |[minimum width=37, fill=gray!60](yOut)| & |[minimum width=9, fill=gray!10] (memory)|\\
};


\path (rec) edge[-] (coordRecSplit);
\path (coordRecSplit) edge[-] (coordRecSplitLeft);
\path (coordRecSplit) edge[-] (coordRecSplitRight);

\path (coordRecSplitRight) edge[-latex] (stateInput);
\path (coordRecSplitLeft) edge[-latex] (symbolInput);

\path (symbolInput) edge[-latex] node[left] {$\nin \times \delta \times \alphabetSize$} (symbolLinEmb);
\path (stateInput) edge[-latex] node[right] {$(\nout + m) \times \delta \times \noutput$} (stateLinEmb);

\path (symbolLinEmb) edge[-latex] node[left] {$\nin \times \hiddenDim$} (symbolPosEmbAdd);
\path (stateLinEmb) edge[-latex] node[right] {$(\nout + m) \times \hiddenDim$} (statePosEmbAdd);
\path (symbolPosEmb) edge[-latex] (symbolPosEmbAdd);
\path (statePosEmb) edge[-latex] (statePosEmbAdd);

\path (symbolPosEmbAdd) edge[-latex] (symbolSelfAttention);
\path (statePosEmbAdd) edge[-latex] (stateSelfAttention);


\path (symbolResidualStart) edge[-latex] (symbolLayerNorm);
\path (symbolLayerNorm) edge[-latex] node[right] {$Q$}(symbolLayerCrossAttn);
\path (symbolLayerCrossAttn) edge[-latex] (symbolResidualAdd);

\path (symbolGeneratorMask) edge[-latex] (symbolIntersectionGeneratorCrossAttn);

\path (symbolSelfAttention) edge[-] (symbolResidualStart);
\path (symbolResidualStart) edge[-] (symbolResidualP1);
\path (symbolResidualP1) edge[-] (symbolResidualP2);
\path (symbolResidualP2) edge[-latex] (symbolResidualAdd);

\path (stateSelfAttention) edge[-] (symbolIntersectionCrossAttnState);

\path (symbolIntersectionCrossAttnState) edge[-] (symbolKeyValueSplit);
\path (symbolKeyValueSplit) edge[-] (symbolKeyPathStart);
\path (symbolKeyPathStart) edge[-latex] node[above] {$K$}(symbolKeyPathEnd);
\path (symbolKeyValueSplit) edge[-] (symbolValuePathStart);
\path (symbolValuePathStart) edge[-latex]  node[below] {$V$}(symbolValuePathEnd);

\path (symbolResidual2Start) edge[-latex] (symbolLayerNorm2);
\path (symbolLayerNorm2) edge[-latex] (symbolFeedForwardLayer);
\path (symbolFeedForwardLayer) edge[-latex] (symbolResidual2Add);

\path (symbolResidualAdd) edge[-] (symbolResidual2Start);
\path (symbolResidual2Start) edge[-] (symbolResidual2P1);
\path (symbolResidual2P1) edge[-] (symbolResidual2P2);
\path (symbolResidual2P2) edge[-latex] (symbolResidual2Add);


\path (stateResidualStart) edge[-latex] (stateLayerNorm);
\path (stateLayerNorm) edge[-latex] node[right] {$Q$}(stateLayerCrossAttn);
\path (stateLayerCrossAttn) edge[-latex] (stateResidualAdd);

\path (stateGeneratorMask) edge[-latex] (stateIntersectionGeneratorCrossAttn);

\path (symbolIntersectionCrossAttnState) edge[-] (stateResidualStart);
\path (stateResidualStart) edge[-] (stateResidualP1);
\path (stateResidualP1) edge[-] (stateResidualP2);
\path (stateResidualP2) edge[-latex] (stateResidualAdd);

\path (symbolResidual2Add) edge[-] (stateIntersectionCrossAttnSymbol);

\path (stateIntersectionCrossAttnSymbol) edge[-] (stateKeyValueSplit);
\path (stateKeyValueSplit) edge[-] (stateKeyPathStart);
\path (stateKeyPathStart) edge[-latex] node[above] {$K$}(stateKeyPathEnd);
\path (stateKeyValueSplit) edge[-] (stateValuePathStart);
\path (stateValuePathStart) edge[-latex]  node[below] {$V$}(stateValuePathEnd);

\path (stateResidual2Start) edge[-latex] (stateLayerNorm2);
\path (stateLayerNorm2) edge[-latex] (stateFeedForwardLayer);
\path (stateFeedForwardLayer) edge[-latex] (stateResidual2Add);

\path (stateResidualAdd) edge[-] (stateResidual2Start);
\path (stateResidual2Start) edge[-] (stateResidual2P1);
\path (stateResidual2P1) edge[-] (stateResidual2P2);
\path (stateResidual2P2) edge[-latex] (stateResidual2Add);


\path (stateResidual2Add) edge [-latex] (stateLayerNormHead);
\path (stateIntersectionCrossAttnSymbol) edge[-latex] (symbolLayerNormHead);

\path (symbolLayerNormHead) edge [-latex] (symbolFeedForwardHead);
\path (stateLayerNormHead) edge [-latex] (stateFeedForwardHead);


\path (symbolFeedForwardHead) edge [-latex] node[right, yshift=-1.5pt] {$\nin$} (symbolFullOutputIntersection);
\path (stateFeedForwardHead) edge [-latex] node[right, yshift=-1.5pt] {$\nout + m$} (stateFullOutputIntersection);

\path (fullOutput) edge [latex'-latex'] node[right] {$\text{BCE}$} (lossIllustration);

\path[draw,decorate,decoration={brace, amplitude=6pt}] ([yshift=3pt, xshift=1pt]yIn.north west) -- ([yshift=3pt, xshift=-1pt]yIn.north east) node[midway,above, yshift=6pt,](yInLabel){$\yin$};
\path[draw,decorate,decoration={brace, amplitude=6pt}] ([yshift=3pt, xshift=1pt]yOut.north west) -- ([yshift=3pt, xshift=-1pt]yOut.north east) node[midway,above, yshift=6pt,]{$\yout$};
\path[draw,decorate,decoration={brace, amplitude=6pt}] ([yshift=3pt, xshift=1pt]memory.north west) -- ([yshift=3pt, xshift=-1pt]memory.north east) node[midway,above, yshift=6pt,]{$\bm{0}^m$};


\begin{pgfonlayer}{background}
\node (ContainerEmbedding) [rectangle, rounded corners= 5mm,  dashed, draw, minimum width=9cm, fit=(symbolPosEmb) (statePosEmb) (rec), label={[shift={(5pt,5pt)}, anchor=south west]south west:\textbf{Embedding Block}}] {};


\node[coordinate] (stateResidual2AddMoreNorth) at ([yshift=2pt]stateResidual2Add.center) {};

\node (ContainerDecoderLayer) [rectangle, rounded corners= 5mm, dashed, draw, minimum width=9cm, fit=(stateResidual2AddMoreNorth) (symbolSelfAttention) (stateResidual2P1) (symbolResidual2P1), label={[shift={(5pt,-5pt)}, anchor=north west]north west:$N\times$\textbf{Decoder Block}}] {};

\node (ContainerLossLayer) [rectangle, rounded corners= 5mm, dashed, draw, minimum width=9cm, fit=(fullOutput) (lossIllustration) (yInLabel), label={[shift={(5pt,-5pt)}, anchor=north west]north west:\textbf{Loss Calculation}}] {};

\end{pgfonlayer}

\end{tikzpicture}
    \caption{Embedding, architecture, and loss calculation for \textit{ConvBCJRFormer}. Visualization inspired by the article~\cite{park2025crossmpt}.}
    \label{fig:convolutionalBCJRFormerArchitecture}
\end{figure}
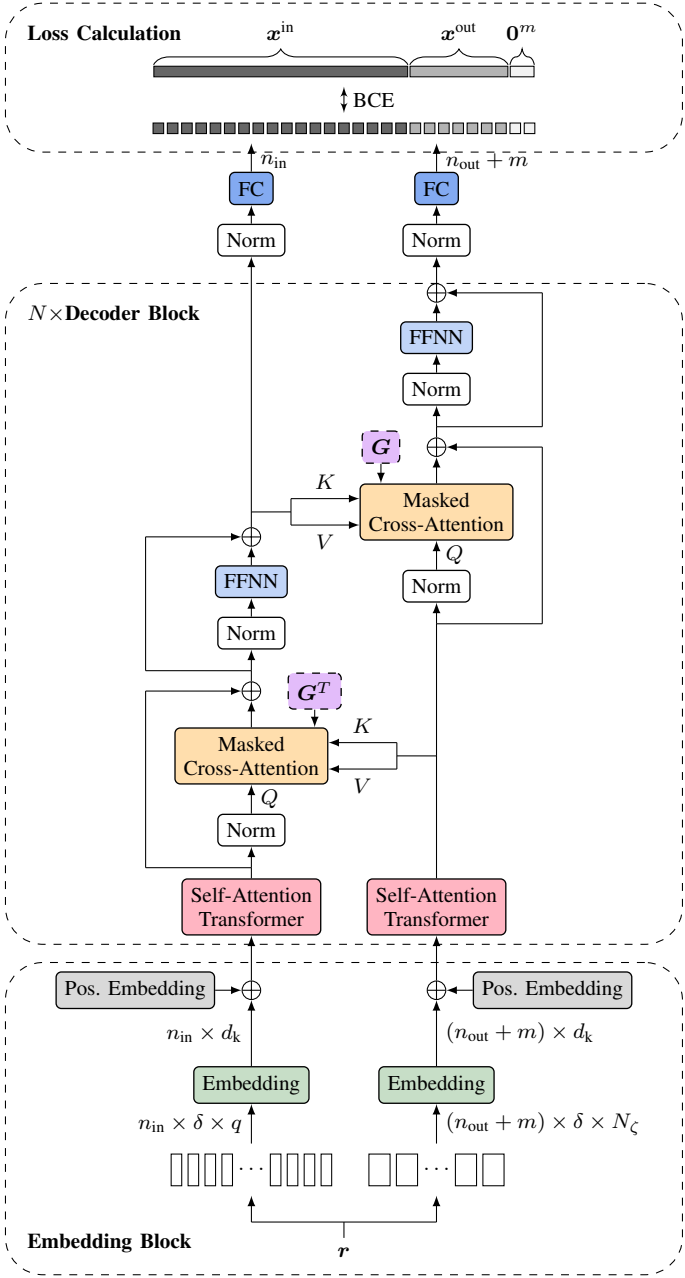
\label{sec:methConvBCJRFormer}
We propose \textit{ConvBCJRFormer}, an extension of \textit{BCJRFormer} that decodes transmissions of a zero-terminated binary convolutional $(\convn, 1, m)$ codeword over the \ac{ids} channel. We denote the input message length by $\nout$ and define the codeword length as $\nin = \convn(\nout + m)$. After encoding, we add a random offset $\offset \in \mathbb{F}_2^{\nin}$.

\subsubsection{Input Construction}

For inner marker codes, each symbol is encoded independently. In these cases, the symbol-wise prior information in the \textit{BCJRFormer} is sufficient. In contrast, encoding a symbol $\youtIx_i$ with a convolutional code depends not only on the symbol itself but also on the encoder's state, which is determined by the previous $m$ symbols. Because the encoded symbols are generated based on the encoder's state, they are not independent. The symbol-wise input used in \textit{BCJRFormer} does not capture the state dependency. Therefore, we extend the input in \textit{ConvBCJRFormer} to include a state-wise representation. 

We use two drift-based sliding windows: one for symbols, denoted by $\bcjrformerInputBit$, and one for states, denoted by $\bcjrformerInputState$. The symbol-based sliding window is identical to the one used for marker codes (see Equation~\eqref{eq:BCJRFormerInput}). The state-based sliding window is a tensor with dimensions $(\nout + m) \times \delta \times \noutput$. Here, $\delta = \dfrom_{\max} - \dfrom_{\min} + 1 $ is the size of the sliding window, and $\noutput = 2^{\convn}$ is the number of possible output sequences $\YStateVal \in \mathbb{F}_2^{\convn}$. These sequences correspond to a convolutional state transition from $\youtIx_{i-1}$ to $\youtIx_{i}$ in the input sequence, where $1 \leq i \leq (\nout + m)$. We denote the output sequence 
$$[\yinIx_{\convn(i-1) + 1}, \yinIx_{\convn(i-1)+ 2}, \ldots, \yinIx_{\convn i}]$$
that arises from this state transition as $\yout_{(i-1) \rightarrow i}$. 

We enumerate all possible output sequences $\YStateVal$ using indices $\YStateValIx = 1, 2, \ldots, \noutput$. Assume that the drift is $\dfrom_j$ before the transmission of $\yout_{(i-1) \rightarrow i}$. We define the scalar element $\bcjrformerInputStateIx_{i, j, \YStateValIx}$ as the joint probability that the transmitted state output $\yout_{(i-1) \rightarrow i}$ equals $\bm{\YStateVal}$ and that the observed output is the sequence $$[\recIx_{\convn(i-1) + \dfrom_j + 1}, \recIx_{\convn(i-1) + \dfrom_j + 2}, \ldots , \recIx_{\convn i + \dfrom_j}],$$ under the simplifying assumption that each transmitted symbol is independently substituted with probability $\psub$ and that there are no insertions or deletions. When we include the additional offset vector $\bm{o}$ (with addition over $\mathbb{F}_2$), the joint probability becomes 
\begin{align*}
    \bcjrformerInputStateIx_{i, j, \YStateValIx}=&P([\yinIx_{\convn(i-1) + 1}, \yinIx_{\convn(i-1)+ 2}, \ldots ,\yinIx_{\convn i}] = \bm{\YStateVal}) \\
    &\cdot \prod_{l=1}^{\convn} F(\yinIx_{\convn(i-1) + l}, \recIx_{\convn(i-1) + l + \dfrom_j} + \offsetIx_{\convn(i-1) + l + \dfrom_j}),
\end{align*} for $1 \leq i \leq (\nout + m)$, $\dfrom_{\min} \leq j \leq \dfrom_{\max}$, and $\YStateValIx = 1, 2, \ldots, \noutput$. 
For simplicity, we assume a balanced convolutional code. This means that, without knowing the encoder's state, all output sequences associated with a state transition $\youtIx_i \rightarrow \youtIx_{i+1}$ are equally likely. Then the prior probabilities become
$$P([\yinIx_{\convn(i-1) + 1}, \yinIx_{\convn(i-1)+ 2}, \ldots, \yinIx_{\convn i}] = \YStateVal) = \frac{1}{\noutput}.$$

The state input representation $\bcjrformerInputState$ links the received sequence to the output of convolutional state transitions. However, assuming that this output is only affected by substitutions is overly simplistic and fails to capture all the information in the received sequence. To overcome this limitation, we also include the symbol-level representation $\bcjrformerInputBit$.  
\subsubsection{Architecture}

For input codeword positions $i$, the encoded symbols $\yinIx_{\convn(i-1)+1}\cc \yinIx_{\convn(i-1)+2} \cc \ldots \cc \yinIx_{\convn i}$ are all generated by the same input symbol $\youtIx_i$ and the same convolutional state. The convolutional code's generator matrix $G$ links these symbols to the state transition that generated them. We incorporate the generator matrix into \mbox{\textit{ConvBCJRFormer}'s} attention mechanism. This integration connects the symbol representation with the state representation and helps the model learn the convolutional code's structure. Figure~\ref{fig:convolutionalBCJRFormerArchitecture} illustrates the modified architecture.  

We first embed the token representations $\bcjrformerInputBit$ and $\bcjrformerInputState$ separately. To synchronize these two input representations, we pass each representation through a separate vanilla transformer that uses unmasked self-attention blocks to capture global information.

To decode the synchronized representations, we use two masked cross-attention layers. We treat the synchronized symbol representation as an abstract representation of the codeword and the synchronized state representation as an abstract representation of the input message, which we aim to recover.
\begin{figure*}[ht]
  \centering %
  \begin{minipage}{.46\textwidth} %
  \begin{tikzpicture}
      \begin{axis}[
          yticklabel style={/pgf/number format/fixed, /pgf/number format/precision=3},
          ybar,   
            ylabel near ticks,
            ylabel shift={-10pt},
          width=\textwidth, 
          xmin=0.00, xmax=0.055,
          ymode=log,
          ymajorgrids=true,
          yminorgrids=true,
          xticklabel style = log ticks with fixed point,
          xtick distance=0.01,
          tick align=inside,
          log origin=infty,
          ymax=0.25,
          ymin=0.01,
          grid=both,
          grid style={line width=.1pt},
          legend pos=south east,
          legend cell align={left},
          xlabel={$\pins = \pdel$},
          ylabel={BER},
          bar width=2pt,              
          enlarge x limits=0,        
            legend image code/.code={
                    \draw [#1] (0cm,-0.1cm) rectangle (3pt,2.5pt); },
            ],

         
        ]
        \addplot[line width=.2pt, blue, fill=blue!50, bar shift=-4.4pt] coordinates { 
        (0.005, 0.013444722493489583)
        (0.01, 0.026736958821614584)
        (0.015, 0.03960759480794271)
        (0.02, 0.05242589314778646) 
        (0.025, 0.06512540181477865) 
        (0.03, 0.07772438049316406) 
        (0.035,0.09030606587727864)
        (0.04, 0.10267847696940104) 
        (0.045,0.11513160705566407) 
        (0.05,0.12736750284830728)
        };
        \addlegendentry{BCJR $N=96$}
        \addplot[line width=.2pt, red, fill=red!50, bar shift=-2.2pt] coordinates { 
        (0.005, 0.013461) 
        (0.01, 0.026641) 
        (0.015, 0.039711)
        (0.02, 0.05268) 
        (0.025, 0.065479)
        (0.03, 0.078267) 
        (0.035,0.090824)
        (0.04, 0.10306) 
        (0.045,0.11572)
        (0.05,0.12825)
        };
        \addlegendentry{BCJRFormer $N=96$}
    
        \addplot[line width=.2pt, dgreen, fill=dgreen!50, bar shift=0pt] coordinates {
          (0.005, 0.01350973690257353) 
          (0.01, 0.02671093510646446) 
          (0.015, 0.03980562097886029)
          (0.02, 0.05269550398284314) 
          (0.025, 0.06561055501302084) 
          (0.03, 0.07839330336626839) 
          (0.035, 0.09135608149509804) 
          (0.04, 0.10425015318627451) 
          (0.045, 0.1173037899241728) 
          (0.05, 0.13060690487132354)
        };
        \addlegendentry{BCJR $N=204$}
    
        \addplot[line width=.2pt, orange, fill=orange!50, bar shift=2.2pt] coordinates {
          (0.005, 0.013516) 
          (0.01, 0.026715) 
          (0.015, 0.039811)
          (0.02, 0.05286) 
          (0.025, 0.065794) 
          (0.03, 0.078986) 
          (0.035, 0.091724) 
          (0.04, 0.1052) 
          (0.045, 0.11843) 
          (0.05, 0.13169)
        };
        \addlegendentry{BCJRFormer $N=204$}
    
      \end{axis}
    \end{tikzpicture}%
  \caption{ \textit{BCJRFormer} performs comparably to \ac{bcjr} for single transmissions at $\psub = 0$. Outer \ac{ldpc} codes of lengths $(96, 48)$ and $(204, 102)$ are concatenated with markers $\marker=001$ inserted every $\markerFreq=6$ bits.}
  \label{fig:bcjrformervsbcjra1}
  \end{minipage}%
  \hfill%
  \begin{minipage}{.46\textwidth}\begin{tikzpicture}
    \begin{axis}[
          yticklabel style={/pgf/number format/fixed, /pgf/number format/precision=3},
          ybar,   
            ylabel near ticks,
            ylabel shift={-10pt},
          xmin=0.00, xmax=0.055,
          ymode=log,
          ymajorgrids=true,
          yminorgrids=true,
          xticklabel style = log ticks with fixed point,
          xtick distance=0.01,
          tick align=inside,
          log origin=infty,
          ymax=0.25,
          ymin=0.01,
          grid=both,
          grid style={line width=.1pt},
          legend pos=south east,
          legend cell align={left},
          xlabel={$\pins = \pdel$},
          ylabel={SER},
          bar width=4pt,              
          enlarge x limits=0,        
            legend image code/.code={
                    \draw [#1] (0cm,-0.1cm) rectangle (3pt,2.5pt); },
            ],

         
        ]
        \addplot [line width=.2pt, blue, fill=blue!50, bar shift=-2.1pt] coordinates {
            (0.005,0.028782730102539063) 
            (0.01,0.045650482177734375) 
            (0.015,0.062319221496582033)  
            (0.02,0.07869071960449218)  
            (0.025,0.09513584136962891) 
            (0.03,0.11120685577392578)   
            (0.035,0.12713218688964845)  
            (0.04,0.14297428131103515) 
            (0.045,0.15853065490722656) 
            (0.05,0.17391563415527345) 
        };

        \addplot [line width=.2pt, red, fill=red!50, bar shift=2.1pt] coordinates {
            (0.005,0.028844) 
            (0.01,0.045757) 
            (0.015,0.062393)  
            (0.02,0.079278)  
            (0.025,0.095433) 
            (0.03,0.1116)   
            (0.035,0.12771)  
            (0.04,0.14339) 
            (0.045,0.15935) 
            (0.05,0.17505) 
        };
         \addlegendentry{BCJR}
         \addlegendentry{BCJRFormer}
    \end{axis}\end{tikzpicture} %
  \caption{For short quaternary codes, \textit{BCJRFormer} achieves error rates comparable to \ac{bcjr}. A protograph $(64, 32)$ \ac{ldpc} code is transmitted with markers \(\marker = 32\) inserted every \(\markerFreq = 6 \) symbols at $\psub = 0.012$.}
  \label{fig:bcjrformer_q4}
  \end{minipage}
\end{figure*}

\noindent\textit{First Cross-Attention Layer:} 
The codeword representation attends to the message representation. Here, we derive the query matrix from the codeword representation and use the key and value matrices from the message representation. For $1 \leq i \leq \nin$ and $1 \leq j \leq (\nout + m)$, we mask the element $(\attQuery\attKey^T)_{ij}$ if the corresponding element in the transposed generator matrix $(\gen^T)_{ij}$ is zero. If the element $(\gen^T)_{ij}$ is non-zero, then the message symbol $j$ was part of the convolutional state that produced codeword symbol $i$. Thus, the generator matrix incorporates information about the encoding process into the attention module.

\noindent\textit{Second Cross-Attention Layer:} 
We repeat the process in the opposite direction. In this layer, the message representation acts as the query and the symbol representation serves as the key and value. This time, we use the untransposed generator matrix $\gen$ as the mask. This design ensures that each token in the message representation attends only to those tokens in the codeword representation corresponding to symbols generated by the state that contains the message symbol. 

Finally, we pass the outputs of the cross-attention layers for the symbol and state representation through distinct linear layers with dimensions $\hiddenDim \times 1$. We then apply a sigmoid function $\sigma$ to produce the prediction probabilities $\predyin$ for the inner codeword $\yin$, $\predyout$ for the outer codeword $\yout$, and $\hat{\bm{0}}^m$ for the final memory state.

\subsection{Training}
We train all models on the binary alphabet using the \ac{bce} loss:
\begin{equation}
\label{eq:BCE}
BCE(\ymodel\cc \bm{x}) = -\frac{1}{n} \sum_{i=1}^{n} x_i\ln(\ymodelIx_i) + (1 - x_i)\ln(1 - \ymodelIx_i),
\end{equation}
which measures the difference between the model predictions $\ymodel$ and the target sequence $\bm{x}$ (i.e., the inner or outer sequence to be decoded). Our training dataset is generated dynamically by generating pseudorandom sequences $\mess \in \alphabet^k$.

For \textit{BCJRFormer}, each generated sequence is first encoded using an \ac{ldpc} outer code and combined with an inner marker code. We then simulate $M$ transmissions over the \ac{ids} channel. When using a binary alphabet, the model is optimized by minimizing the \ac{bce} loss between the inner decoder's predictions $\predyin$ and the corresponding inner codeword bits $\yin$. For non-binary alphabets, we replace the \ac{bce} loss with multi-class cross entropy.

The outer \ac{ecct} decoder is designed to learn the inner decoder's multiplicative noise. This noise is defined as $$\bm{z} = \ybipolar \cdot \bipolar(\yout),$$ and the decoder is trained by minimizing the loss $\text{BCE}(\bm{\hat{z}}, \bm{\bin(z)})$, where $\bm{\hat{z}}$ is the \ac{ecct}'s prediction. The final predicted outer codeword is then obtained by $$\predyout = \bin(-\bm{\hat{z}} \cdot \text{sign}(\ybipolar)).$$

For the convolutional decoder \textit{ConvBCJRFormer}, we encode each generated sequence $\mess \in \mathbb{F}_2^k$ using a convolutional code. A new pseudorandom offset $\offset$ is generated for each transmission. We optimize the \ac{bce} loss computed between the concatenated model predictions \,---\, $\predyin$, $\predyout$, and $\hat{\bm{0}}^m$\,---\, and the concatenated target sequences \,---\,$\yin$, $\yout$ and $\bm{0}^m$. 

\subsection{Complexity}
We analyze the complexity of jointly decoding multiple noisy copies of a codeword using the \ac{bcjr} algorithm and \textit{BCJRFormer} to justify our deep learning-based approach. 

As in Section~\ref{sec:BCJRFormer}, we restrict the drift states of the \ac{bcjr} decoder at each time step to the range from \(d_{\min}\) to \(d_{\max}\), where the total number of drift states is defined as $\delta = \dfrom_{\max} - \dfrom_{\min} + 1$.
When decoding \(M\) copies jointly, the decoder must track \(\delta^M\) distinct drift states. Each drift state has \(\kappa = I_{\max} + 2\) possible transitions, where \(I_{\max}\) is the maximum number of insertions allowed per symbol and the additional $2$ accounts for a deletion or a transmission. As a result, the overall complexity of the \ac{bcjr} decoding process is \(O(\nin (\kappa \delta)^M)\), where \(\nin\) denotes the input sequence length~\cite{maaroufConcatenatedCodesMultiple2023a}.

In contrast, \textit{BCJRFormer} scales only quadratically with the number of copies $M$. The conversion of the received sequences into the model's input (as detailed in Equation~\eqref{eq:BCJRFormerInput}) and the subsequent embedding have a combined complexity of $O(M\nin\delta\alphabetSize \hiddenDim)$. Each transformer block incurs quadratic complexity, specifically \(O\left(M^2 \nin^2 \Ndec \hiddenDim\right)\), where \(\Ndec\) is the number of attention layers~\cite{vaswaniAttentionAllYou2017a}.

\section{Experimental Results}

We show that our proposed \textit{BCJRFormer} model decodes multiple noisy copies of transmitted sequences with error rates comparable to joint \ac{bcjr} decoding. Moreover, \textit{BCJRFormer} scales to number of sequences that are computationally infeasible for the \ac{bcjr} algorithm. We also demonstrate that combining \textit{BCJRFormer} with \ac{ecct} yields better performance than using either the \ac{bcjr} algorithm or \ac{bp}. Finally, we show that \textit{ConvBCJRFormer} can jointly synchronize and decode convolutional codes with only minor trade-offs compared to the \ac{bcjr} algorithm. 

\subsection{Setup}
We train all neural decoders with the Adam optimizer~\cite{adamAMethodForStoachsticOptimization}. Each \textit{BCJRFormer} model has a hidden dimension of $96$, six attention layers, and eight attention heads per layer, for a total of about 700{,}000 parameters. We train at a batch size of $256$ for 160{,}000 iterations without dropout, and we use a cosine decay learning rate schedule that starts at $\expnumber{2.5}{-4}$ and decays to $\expnumber{2.5}{-5}$ after a warmup of 20{,}000 iterations.

All \ac{ecct} models have a hidden dimension of $128$, with eight attention layers and eight attention heads per layer, for a total of about 1.6 million parameters. We train the models at a batch size of $1024$ for 120{,}000 iterations using a constant learning rate of $\expnumber{1}{-4}$ and no dropout.

For inner decoders we restrict the drift states $d_{\min}$ and $d_{\max}$ to $\pm 5 \sqrt{\nin \frac{\pdel}{1 - \pdel}}$ and fix the maximum number of insertions per transmitted symbol to $I_{\max} = 2$~\cite{daveyReliableCommunicationChannels2001}. When we use \ac{bp} for outer decoding, we run $50$ iterations to ensure convergence. 

We measure performance using the \acf{ber} for binary transmissions and the \acf{ser} for quaternary transmissions. The \acl{ber} and \acl{ser} are defined as the ratio of incorrectly decoded bits or symbols to the total number of transmitted bits or symbols, respectively. We average error rates over 409{,}600 randomly generated codewords, unless noted otherwise.

\subsection{BCJRFormer for Single Marker Codewords}
\label{sec:ExperimentBCJRDecoderSingle}

We compare \textit{BCJRFormer} and the \ac{bcjr} algorithm for inner decoding of single binary sequences using a short \((96, 48)\) \ac{ldpc} code~\cite{channelcodes} and a longer \((204, 102)\) \ac{ldpc} code~\cite{mackayEncyclopediaSparseGraph}. For the short code, we insert the marker sequence \(\marker = 001\) every \(\markerFreq = 6\) bits, and for the longer code, every \(\markerFreq = 7\) bits. This yields inner codeword lengths of $144$ and $291$ with rates of \(2/3\) and approximately $0.7$, respectively. We trained several models over a range of deletion probabilities (set equal to the insertion probability) while fixing the substitution probability at \(\psub = 0\). The \ac{ber} results in Figure~\ref{fig:bcjrformervsbcjra1} show that \textit{BCJRFormer} closely matches the performance of the \ac{bcjr} algorithm for both code lengths across all insertion and deletion rates.

\subsection{BCJRFormer for Quaternary Marker Codewords}
\label{sec:ExperimentBCJRDecoderSingleMarkerQ4}

We also demonstrate that \textit{BCJRFormer} performs well for non-binary codes. We use an outer \((64, 32)\) quaternary protograph \ac{ldpc} code, as proposed in the paper~\cite{maaroufConcatenatedCodesMultiple2023a}. The shortest cycle in the corresponding Tanner graph has a length of eight. The weights in the code's parity-check matrix were set uniformly at random once and then fixed for all experiments. For more details on the construction and the corresponding protograph, see Appendix~\ref{sec:appQuartProtographConstruction}.

We insert the marker sequence $\marker = 32$ every $\markerFreq = 6$ symbols, resulting in an inner codeword length of $\nin = 84$. We vary the insertion and deletion probabilities (with $\pins = \pdel$) and fix the channel's substitution probability at $\psub = 0.012$. We compare the \ac{ser} of the trained models with that of the \ac{bcjr} algorithm. Figure~\ref{fig:bcjrformer_q4} shows that \textit{BCJRFormer} remains competitive across all channel configurations. For a more detailed analysis of the symbol-wise differences between the two decoders, see Appendix~\ref{sec:appBCJRFormerQuartDistribution}.

\begin{figure*}[t]
  \centering %
  \begin{minipage}{.46\textwidth}
    \begin{tikzpicture}
    \begin{axis}[
        ymode=log,
        xmin=0.005, xmax=0.099,
        xticklabel style = {/pgf/number format/fixed, /pgf/number format/precision=3},
        xtick distance=0.02,
        grid = both,
        grid style = {line width=.1pt},
        legend pos=south east,
        legend cell align={left},
        xlabel = {$\pins=\pdel$},
        ylabel = {$\text{BER}$},
        ]

        \addplot [mark=square, mark size=1.5pt, color=blue] coordinates {
        (0.01, 0.015905736287434896)
        (0.02, 0.02536626180013021) 
        (0.03, 0.038973185221354165) 
        (0.04, 0.055336863199869794)
        (0.05,0.0737860107421875)
        (0.06, 0.09353876749674479)
        (0.07, 0.11411542256673177)
        (0.08, 0.13538459777832032)
        };
        \addlegendentry{BCJR}

        
        \addplot [mark=square, mark size=1.5pt, color=red] coordinates {
            (0.01,0.015893) 
            (0.02,0.026096) 
            (0.03,0.04022)  
            (0.04,0.057613)  
            (0.05,0.077038) 
            (0.06,0.097053)   
            (0.07,0.11969)  
            (0.08,0.14178) 
        } node [right=3pt, yshift=2pt, color=black] {\scriptsize$M = 2$};

        \addlegendentry{BCJRFormer}
        \addplot [mark=square, mark size=1.5pt, color=blue] coordinates {
        (0.01, 0.0020398457845052084)
        (0.02, 0.0051534016927083336) 
        (0.03, 0.010960133870442708) 
        (0.04, 0.01949793497721354)
        (0.05,0.030600738525390626)
        (0.06, 0.04468994140625)
        (0.07, 0.061186726888020834)
        (0.08, 0.08055801391601562)
        };
        
        \addplot [mark=square, mark size=1.5pt, color=red] coordinates {
            (0.01,0.002015) 
            (0.02,0.0055537) 
            (0.03,0.0121)  
            (0.04,0.021623)  
            (0.05,0.034531) 
            (0.06,0.049948)   
            (0.07,0.066847)  
            (0.08,0.088706) 
        } node [right = 3pt, yshift=1.5pt, color=black] {\scriptsize$M = 3$};
        
        
        \addplot [mark=square, mark size=1.5pt, color=red] coordinates {
            (0.01,0.0006719) 
            (0.02,0.0017107) 
            (0.03,0.004263)  
            (0.04,0.0091625)  
            (0.05,0.016446) 
            (0.06,0.027271)   
            (0.07,0.041486)  
            (0.08,0.057871) 
        } node [right=3pt, color=black] {\scriptsize$M = 4$};


        \addplot [mark=square, mark size=1.5pt, color=red] coordinates {
            (0.01,0.00013855) 
            (0.02,0.00050873) 
            (0.03,0.0015559)  
            (0.04,0.0037613)  
            (0.05,0.0080445) 
            (0.06,0.014255)   
            (0.07,0.025077)  
            (0.08,0.038179) 
        } node [right=3pt, yshift=-1.5pt, color=black] {\scriptsize$M = 5$};


        \addplot [mark=square, mark size=1.5pt, color=red] coordinates {
            (0.01,0.00001147) 
            (0.02,0.000018463) 
            (0.03,0.000049667)  
            (0.04,0.00011854)  
            (0.05,0.00037768) 
            (0.06,0.00092603)   
            (0.07,0.0021341)  
            (0.08,0.0048321) 
        } node [right=3pt, color=black] {\scriptsize$M = 10$};
        
    \end{axis}
  \end{tikzpicture} %
    \caption{Joint inner decoding performance of \textit{BCJRFormer} scales similarly to that of the \ac{bcjr} algorithm across various cluster sizes \(M\). A \((96, 48)\) \ac{ldpc} outer code is transmitted with markers \(\marker = 001\) inserted every \(\markerFreq = 6\) bits, at \(\psub = 0.012\).}
    \label{fig:bcjrformervsbcjrbiga}
  \end{minipage}%
  \hfill %
  \begin{minipage}{.46\textwidth} 
    \begin{tikzpicture}
    \begin{axis}[
        xmin=0.005, xmax=0.099,
        xticklabel style = {/pgf/number format/fixed, /pgf/number format/precision=3},
        xtick distance=0.02,
        ymode=log,
        grid = both,
        grid style = {line width=.1pt},
        legend pos=south east,
        legend cell align={left},
        xlabel = {$\pins=\pdel$},
        ylabel = {$\text{BER}$},
        ]

         \addlegendentry{BCJRFormer}
         \addlegendentry{Dynamic BCJRFormer}


        \addplot [mark=square, mark size=1.5pt, color=blue] coordinates {
            (0.01,0.040609) 
            (0.02,0.068502) 
            (0.03,0.094556)  
            (0.04,0.11949)  
            (0.05,0.14426) 
            (0.06,0.16769)   
            (0.07,0.19097)  
            (0.08,0.21291) 
        } node [right=3pt, color=black] {\scriptsize$M = 1$};


        \addplot [mark=square, mark size=1.5pt, color=red] coordinates {
            (0.01,0.04080335105303675) 
            (0.02,0.06867175032617524) 
            (0.03,0.09495181007077917)  
            (0.04,0.12032539650332183)  
            (0.05,0.1449454294424504) 
            (0.06,0.16884705077391118)   
            (0.07,0.19214666226878763)  
            (0.08,0.2143152667582035) 
        };


        \addplot [mark=square, mark size=1.5pt, color=blue] coordinates {
            (0.01,0.015893) 
            (0.02,0.026096) 
            (0.03,0.04022)  
            (0.04,0.057613)  
            (0.05,0.077038) 
            (0.06,0.097053)   
            (0.07,0.11969)  
            (0.08,0.14178) 
        } node [right=3pt, color=black] {\scriptsize$M = 2$};


        \addplot [mark=square, mark size=1.5pt, color=red] coordinates {
            (0.01,0.01632209830917418) 
            (0.02,0.02651997947250493) 
            (0.03,0.04097508871811442)  
            (0.04,0.05878545263549313)  
            (0.05,0.0779997786716558) 
            (0.06,0.09899233747972175)   
            (0.07,0.12116391800111159)  
            (0.08,0.14375885498244315) 
        };

        \addplot [mark=square, mark size=1.5pt, color=blue] coordinates {
            (0.01,0.002015) 
            (0.02,0.0055537) 
            (0.03,0.0121)  
            (0.04,0.021623)  
            (0.05,0.034531) 
            (0.06,0.049948)   
            (0.07,0.066847)  
            (0.08,0.088706) 
        } node [right=3pt, color=black] {\scriptsize$M = 3$};
        

        \addplot [mark=square, mark size=1.5pt, color=red] coordinates {
            (0.01,0.0023319753705800393) 
            (0.02,0.0059244284614396745) 
            (0.03,0.012543411572405603)  
            (0.04,0.022754034040845)  
            (0.05,0.03540250265330542) 
            (0.06,0.05092089458019473)   
            (0.07,0.06960914847441017)  
            (0.08,0.09060961658833548) 
        };

        \addplot [mark=square, mark size=1.5pt, color=blue] coordinates {
                    (0.01,0.0006719) 
                    (0.02,0.0017107) 
                    (0.03,0.004263)  
                    (0.04,0.0091625)  
                    (0.05,0.016446) 
                    (0.06,0.027271)   
                    (0.07,0.041486)  
                    (0.08,0.057871) 
        } node [right=3pt, color=black] {\scriptsize$M = 4$};

        \addplot [mark=square, mark size=1.5pt, color=red] coordinates {
            (0.01,0.0007393392133644739) 
            (0.02,0.0017986806776025333) 
            (0.03,0.004518432755139657)  
            (0.04,0.009727630918059732)  
            (0.05,0.017482656429056078) 
            (0.06,0.02790608788956888)   
            (0.07,0.04211670045624487)  
            (0.08,0.06005127107258886) 
        };

        \addplot [mark=square, mark size=1.5pt, color=blue] coordinates {
            (0.01,0.00013855) 
            (0.02,0.00050873) 
            (0.03,0.0015559)  
            (0.04,0.0037613)  
            (0.05,0.0080445) 
            (0.06,0.014255)   
            (0.07,0.025077)  
            (0.08,0.038179) 
        } node [right=3pt, color=black] {\scriptsize$M = 5$};
        

        \addplot [mark=square, mark size=1.5pt, color=red] coordinates {
            (0.01,0.00015586853385912036) 
            (0.02,0.0005132548159076577) 
            (0.03,0.001658223518752493)  
            (0.04,0.004266484706604388)  
            (0.05,0.00865460741755669) 
            (0.06,0.015558599291834981)   
            (0.07,0.025912196455756202)  
            (0.08,0.039996974917594345) 
        };     
        
    \end{axis}
  \end{tikzpicture} %
  \caption{\textit{BCJRFormer} models trained with various cluster sizes perform comparably to those trained on fixed cluster sizes \(M\). Markers \(\marker = 001\) are inserted every \(\markerFreq = 6\) bits into a $(96, 48)$ \ac{ldpc} outer code, and transmitted at \(\psub = 0.012\).}
  \label{fig:bcjrformervsbcjrdynamicseq}
    \end{minipage} 
\end{figure*}

\begin{figure*}[t]
  \begin{minipage}{.46\textwidth}
    \begin{tikzpicture}
        \begin{axis}[
            xmin=0.0,   xmax=0.055,
            width=\textwidth,
            scaled ticks=true,
            yticklabel style = {/pgf/number format/fixed, /pgf/number format/precision=3},
            xticklabel style = log ticks with fixed point,
            xtick distance=0.01,
            ymode=log,
            grid = both,
            grid style = {line width=.1pt},
            legend pos=south east,
            legend cell align={left},
            tick align=inside,
            xlabel = {$\pins=\pdel$},
            ylabel = {BER},
            ]
        \addplot [mark=square, mark options={solid}, mark size=1.5pt,color=blue] 
        coordinates {
            (0.005, 0.0025617)
            (0.01, 0.0069188)
            (0.015, 0.012154)
            (0.02, 0.021845)
            (0.025, 0.034011)
            (0.03, 0.049067)
            (0.035,0.065758)
            (0.04, 0.083304) 
            (0.045,0.10181)
            (0.05,0.11971)
         };
        \addlegendentry{ECCT - Marker }

        \addplot [mark=square,  mark options={solid}, mark size=1.5pt, color=red] coordinates { 
            (0.005, 0.006616439819335938) 
            (0.01, 0.013698399861653646) 
            (0.015, 0.02289576212565104)
            (0.02, 0.034660720825195314)
            (0.025,0.04897631327311198)
            (0.03, 0.06457809448242187)
            (0.035,0.08133433024088542) 
            (0.04, 0.09816968282063802) 
            (0.045,0.11507397969563803)
            (0.05,0.13100939432779948)
        };
        \addlegendentry{BP - Conv}
        \addplot [dashed,mark=square, mark options={solid},  mark size=1.5pt,color=blue] coordinates {
            (0.005, 0.0015917) 
            (0.01, 0.0011578) 
            (0.015, 0.0011021)
            (0.02, 0.0022714)
            (0.025, 0.0039464)
            (0.03, 0.0078504)
            (0.035,0.014509)
            (0.04, 0.023818) 
            (0.045,0.037381)
            (0.05,0.054488)
         };
        \addlegendentry{ECCT - Conv. }
        
        \addplot [dashed,mark=square, mark options={solid}, mark size=1.5pt, color=red] coordinates {
            (0.005, 0.0011722056070963543)
            (0.01, 0.001597900390625)
            (0.015, 0.0028933207194010418)
            (0.02, 0.005740712483723959) 
            (0.025,0.010032119750976563)
            (0.03, 0.016975428263346356)
            (0.035,0.026866302490234376)
            (0.04, 0.040261637369791665) 
            (0.045,0.056417236328125)
            (0.05,0.07596430460611979)
         };
        \addlegendentry{BP - Conv.}

        \end{axis}
    \end{tikzpicture}%
    \caption{The outer \ac{ecct} decoder outperforms \acf{bp} for a \((96, 48)\) \ac{ldpc} outer code at $\psub = 0.012$, as shown for a rate-$1/2$ convolutional code (\( g = [5, 7]_8 \)), and a $001$ marker code with \(\markerFreq = 6\).}
    \label{fig:experiment_outer_ecct_n96_k48}
  \end{minipage}%
  \hfill%
  \begin{minipage}{.46\textwidth}
    \begin{tikzpicture}
      \begin{axis}[
          ybar,   
          xmin=0.00, xmax=0.055,
          xtick distance=0.01,
          ymode=log,
          log origin=infty,
          ymax=0.155,
          xtick align=inside,
          grid=both,
          ylabel near ticks,
          ylabel shift={-10pt},
          grid style={line width=.1pt},
          legend pos=south east,
          legend cell align={left},
          xticklabel style = log ticks with fixed point,
          xlabel={$\pins = \pdel$},
          ylabel={$ \text{BER} $},
          bar width=2pt,              
          enlarge x limits=0,        
            legend image code/.code={
                    \draw [#1] (0cm,-0.1cm) rectangle (3pt,2.5pt); },
            ]
          
         
        ]


        \addplot [line width=.2pt, blue, fill=blue!50, bar shift=-4.4pt] coordinates { 
            (0.005, 0.00018829) 
            (0.01, 0.0010532)  
            (0.015, 0.003039)  
            (0.02, 0.0074604)   
            (0.025, 0.014794)  
            (0.03, 0.024072)  
            (0.035,0.039136)  
            (0.04, 0.05459) 
            (0.045,0.072515) 
            (0.05,0.091268) 
        };
        \addlegendentry{BCJRFormer + ECCT}

        
        \addplot [line width=.2pt, red, fill=red!50, bar shift=-2.2pt] coordinates{ 
            (0.005, 0.00071014) 
            (0.01, 0.0022634)  
            (0.015, 0.0033396)  
            (0.02, 0.0074259)  
            (0.025, 0.014275)  
            (0.03, 0.024583)  
            (0.035,0.037592)  
            (0.04, 0.053496) 
            (0.045,0.071681) 
            (0.05,0.090808) 
        };
        \addlegendentry{BCJR + ECCT}
        
        \addplot [line width=.2pt, dgreen, fill=dgreen!50, bar shift=0pt] coordinates { 
            (0.005, 0.0018696594238281259) 
            (0.01, 0.004776280721028644)  
            (0.015, 0.00948842366536459)  
            (0.02, 0.017035369873046873)  
            (0.025, 0.02741559346516925)
            (0.03, 0.04031443277994788) 
            (0.035,0.05561790466308592)  
            (0.04, 0.07258878072102865) 
            (0.045,0.08980473836263031) 
            (0.05,0.10747568766276042) 
        };
        \addlegendentry{BCJRFormer + BP}
        
        \addplot [line width=.2pt, orange, fill=orange!50, bar shift=2.2pt] coordinates { 
            (0.005, 0.002160059611002604) 
            (0.01, 0.005485254923502604)  
            (0.015, 0.009573135375976562)  
            (0.02, 0.017008590698242187)  
            (0.025, 0.02712163289388021)  
            (0.03, 0.040102335611979165)  
            (0.035,0.055196253458658855)  
            (0.04, 0.07171783447265626) 
            (0.045,0.08895790100097656) 
            (0.05,0.10665359497070312) 
        };
        \addlegendentry{BCJR + BP}
        \end{axis}
    \end{tikzpicture}
    \caption{Transformer-based decoder combinations achieve lower error rates than iterative algorithms, as shown at \(\psub = 0.0\) with a \((96, 48)\) \ac{ldpc} code and markers \(\marker = 001\) inserted every \(\markerFreq = 6\) bits.}
    \label{fig:experiment_e2e_separate}
  \end{minipage}
\end{figure*}

\subsection{BCJRFormer for Joint Inner Decoding}
Next, we evaluate \textit{BCJRFormer} for jointly decoding \(M\) noisy copies of a codeword. We train separate inner decoders for different values of \(M\) and for varying deletion and insertion probabilities \(\pdel = \pins\) with substitution probability \(\psub = 0.012\), and we compare their error rates with those of the \ac{bcjr} algorithm. For \(M = 3\), we evaluate the \ac{bcjr} algorithm's error rates over only 40{,}960 samples because its complexity grows exponentially. As shown in Figure~\ref{fig:bcjrformervsbcjrbiga}, the error rates of \textit{BCJRFormer} scale in a manner similar to those of the \ac{bcjr} algorithm as the number of received copies increases. For values \(M > 3\), \textit{BCJRFormer} attains decreasing \acp{ber}, whereas the \ac{bcjr} algorithm becomes computationally impractical.

In \ac{dna} data storage, the number of received sequences (denoted as \(M_k\)) varies for each transmitted sequence, which makes training models for fixed cluster sizes inefficient. We demonstrate that a single \textit{BCJRFormer} model can decode sequences when the number of copies varies between $M_{\min}$ and $M_{\max}$. 

To handle varying \(M_k\), we pad the concatenated input\,---\,which originally has dimensions $\nin M_k \times \delta \times q$ \,---\, with a dedicated token so that it has dimensions \(\nin M_{\max} \times \delta \times q\), and we mask the \(M_{\max} - M_k\) padded columns in both the attention mechanism~\eqref{eq:attention} and the mean aggregation~\eqref{eq:mean_dim_reduction}. 

We train dynamic models across a range of deletion and insertion probabilities ($\pins = \pdel$) with substitution probability $\psub$ by sampling $M_k$ uniformly at random from the range $[M_{\min}, M_{\max}]$. We evaluate the dynamically trained model by comparing it to models trained on the same channel configuration with a fixed \(M_k\). As shown in Figure~\ref{fig:bcjrformervsbcjrdynamicseq}, the dynamic model performs on par with models trained using a fixed $M_k$ and even attains lower \acp{ber} in some scenarios.

\subsection{ECCT for Outer Decoding}

\label{sec:ExperimentECCTOuterDecoding}
We compare the \ac{ecct} decoder with the \ac{bp} decoder using the short \((96, 48)\) \ac{ldpc} code from Section~\ref{sec:ExperimentBCJRDecoderSingle}. We consider two inner code constructions. First, we concatenate the \ac{ldpc} code with a $(2, 1, 2)$ zero-terminated convolutional code with generator polynomials \(g = (5, 7)_8\). The codeword is combined with a pseudorandom offset which varies with each transmission. Secondly, we insert markers, $\marker = 001$, at a fixed interval $\markerFreq$. We train models using a range of deletion and insertion probabilities (with \(\pins = \pdel\)) and a fixed substitution probability $\psub = 0.012$ using the a posteriori probability approximations from an inner \ac{bcjr} decoder. Figure~\ref{fig:experiment_outer_ecct_n96_k48} shows that the \ac{ecct} decoder outperforms \ac{bp} decoding across most channel configurations. We note that the performance gap between \ac{ecct} and \ac{bp} widens in the low probability domain. However, when using inner convolutional codes at extremely low deletion and insertion probabilities, the performance of \ac{ecct} deteriorates.

\subsection{End-to-End Transformer Decoding}

We demonstrate that an end-to-end transformer pipeline\,---\,using \textit{BCJRFormer} as the inner decoder and \ac{ecct} as the outer decoder\,---\,outperforms pipelines that use either the \ac{bcjr} algorithm or \ac{bp} as decoders. We reuse the \textit{BCJRFormer} models from Section~\ref{sec:ExperimentBCJRDecoderSingle} and train each \ac{ecct} outer decoder on the outputs produced by the inner decoder for a channel configuration with equal insertion and deletion rates ($\pins = \pdel$) and a substitution probability of $\psub = 0.012$. Figure~\ref{fig:experiment_e2e_separate} compares various combinations of inner and outer decoders. We observe that the \ac{ecct} yields lower error rates than \ac{bp} decoding across all probability ranges. In the high probability domain, the pipeline using the iterative \ac{bcjr} inner decoder slightly outperforms the pipeline using \textit{BCJRFormer}. In the low probability domain, all decoder combinations that use \textit{BCJRFormer} as the inner decoder significantly outperform those that use the \ac{bcjr} algorithm. These results shows that transformer-based pipelines can outperform iterative approaches in concatenated coding.

\subsection{ConvBCJRFormer for Convolutional Decoding}
\label{sec:ExpConvBCJRFormer}
We hypothesize that a major benefit of neural decoders is their ability to incorporate outer code information during synchronization which is computationally infeasible within the \ac{bcjr} algorithm when applied to general linear codes~\cite{bahlOptimalDecodingLinear1974}. In a first step toward incorporating linear code information, we consider the joint synchronization and decoding of linear convolutional codes. 

We compare the error rates of our proposed \textit{ConvBCJRFormer} decoder with those of the \ac{bcjr} algorithm. We construct the input as described in Section~\ref{sec:methConvBCJRFormer}, setting the drift window size $\delta$ of the state inputs equal to that of the symbol inputs. The \mbox{\textit{ConvBCJRFormer}} models have a hidden dimension of $\hiddenDim = 96$ and consist of four decoder blocks. Both the symbol and state self-attention transformers have three attention layers, followed by a single cross-attention block. All attention layers have six heads. In total, each model has approximately $3.6$ million learnable parameters. 

We train each \textit{ConvBCJRFormer} model for 480{,}000 iterations with a batch size of $512$ and no dropout. We use a cosine decay learning rate schedule that starts at $\expnumber{2.75}{-4}$ and decays to $\expnumber{2.75}{-5}$ after a 20{,}000-iteration warmup. We encode a \ac{ldpc} $(96, 48)$ outer codeword using a rate-$1/2$ zero-terminated $[5, 7]_8$ convolutional code, as introduced in Section~\ref{sec:methConvBCJRFormer}. The resulting convolutional codeword has a length of $\nin = 196$. We fix the \ac{ids} channel's substitution rate at $\psub = 0.012$ and train the models over a range of equal insertion and deletion rates ($\pins = \pdel$). 

\begin{figure}[t]
    \centering
    \begin{tikzpicture}
        \begin{axis}[
                    xmin=0.0,   
                    xmax=0.055,
                    scaled ticks=true,
                    yticklabel style = {/pgf/number format/fixed, /pgf/number format/precision=3},
                    xticklabel style = log ticks with fixed point,
                    xtick distance=0.01,
                    ymode=log,
                    grid = both,
                    grid style = {line width=.1pt},
                    legend pos=south east,
                    legend cell align={left},
                    tick align=inside,
                    xlabel = {$\pins=\pdel$},
                    ylabel = {BER},
                    ]
            \addplot [mark=square,  mark options={solid}, mark size=1.5pt, color=blue] coordinates {
                (0.005,0.003501459757486979) 
                (0.01,0.008185933430989584) 
                (0.015,0.01450347900390625)  
                (0.02,0.022513936360677084)  
                (0.025,0.032356516520182295) 
                (0.03,0.04393285115559896)   
                (0.035,0.057135365804036456)  
                (0.04,0.07177179972330729) 
                (0.045,0.08770693461100261) 
                (0.05,0.10490519205729167) 
            };
    
            \addplot [mark=square,  mark options={solid}, mark size=1.5pt, color=red] coordinates {
                (0.005,0.0037358) 
                (0.01,0.0098614) 
                (0.015,0.017628)  
                (0.02,0.027661)  
                (0.025,0.041249) 
                (0.03,0.053734)   
                (0.035,0.070144)  
                (0.04,0.086839) 
                (0.045,0.10761) 
                (0.05,0.12837) 
            };
            \addplot [dashed, mark=square,  mark options={solid}, mark size=1.5pt, color=blue] coordinates {
                (0.005,0.0011664835611979168) 
                (0.01,0.00292022705078125) 
                (0.015,0.005739491780598958)  
                (0.02,0.009985555013020833)  
                (0.025,0.015766881306966147) 
                (0.03,0.02329432169596354)   
                (0.035,0.03266047159830729)  
                (0.04,0.04392194112141927) 
                (0.045,0.057304941813151045) 
                (0.05,0.07221783955891926) 
            };
    
            \addplot [dashed, mark=square,  mark options={solid}, mark size=1.5pt, color=red] coordinates {
                (0.005,0.0012084) 
                (0.01,0.003556) 
                (0.015,0.0079335)  
                (0.02,0.013748)  
                (0.025,0.022155) 
                (0.03,0.031114)   
                (0.035,0.043672)  
                (0.04,0.057235) 
                (0.045,0.074556) 
                (0.05,0.092399) 
            };
             \addlegendentry{BCJR; $\psub=0.012$}
             \addlegendentry{ConvBCJRFormer; $\psub=0.012$}             \addlegendentry{BCJR; $\psub=0.0$}
             \addlegendentry{ConvBCJRFormer; $\psub=0.0$}
        \end{axis}\end{tikzpicture} %
      \caption{Error rates of \textit{ConvBCJRFormer} are marginally higher than those of the \ac{bcjr} algorithm, shown for a rate-$1/2$ convolutional code with $g = [5, 7]_8$ and an input codeword length of $\nout = 96$.}
      \label{fig:ConvBCJRFormer}
\end{figure}
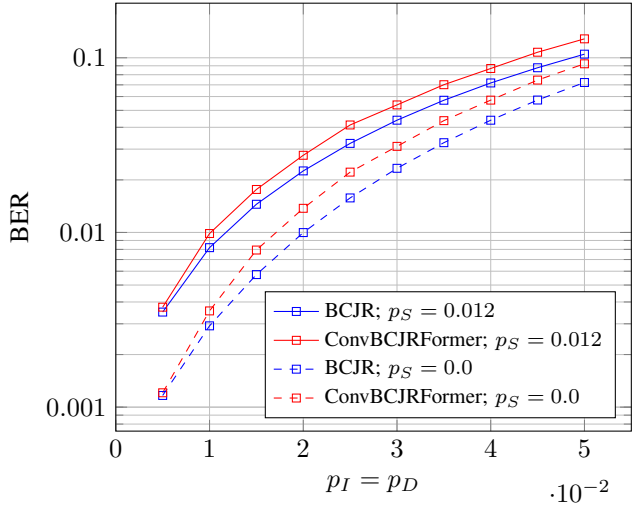

In Figure~\ref{fig:ConvBCJRFormer}, we compare the difference in \acp{ber} between the decoded outer codeword $\predyout$ and the outer codeword $\yout$ for both the \ac{bcjr} algorithm and the \mbox{\textit{ConvBCJRFormer}} decoder. While error rates of our proposed model are slightly higher than those of the \ac{bcjr} algorithm, we observe that our proposed architecture successfully learns the convolutional code structure. This becomes more clear when comparing our model with models that do not employ any attention masking, as we demonstrate in Appendix~\ref{sec:appTrainConvergenceOfConvBCJRFormer}.

\section{Conclusion}
In this work, we propose a high-performance and efficient two-step transformer-based decoding approach for handling multiple noisy copies of a codeword transmitted over the \ac{ids} channel.

By replacing the \ac{bcjr} algorithm with \textit{BCJRFormer} as the inner decoder and substituting \ac{bp} with \ac{ecct} as the outer decoder, our end-to-end pipeline achieves lower error rates compared to iterative algorithms while efficiently scaling to cluster sizes that are not feasible with the \ac{bcjr} algorithm. Our study is currently limited to randomly generated sequences. A promising direction for future study is to apply and fine-tune our methodology on real \ac{dna} traces. In this context, it would be valuable to compare the performance of \textit{BCJRFormer} for coded multiple sequence alignment with other methods, such as those proposed in the papers~\cite{srinivasavaradhanTrellisBMACoded2021, maaroufConcatenatedCodesMultiple2023a}. \\
We further introduced \textit{ConvBCJRFormer}, a transformer-based architecture that jointly synchronizes and decodes convolutional codes. A natural extension for future work is to incorporate the structure of more general linear codes into the inner decoding process. This approach could improve synchronization and overcome the computational limitations of the \ac{bcjr} algorithm.   
\newpage
\IEEEtriggeratref{18}
\bibliography{definitions,bibliofile}

\newpage

\onecolumn
\appendices
\begin{figure}[h]
    \centering
    \subfloat[Layer 1]{
    \label{subfig:AttExampleLayer0}
    \begin{tikzpicture}[scale=.9]
    \begin{axis}[
    tick align=outside,
    tick pos=both,
    axis line style={draw=none} ,
    xtick={6,7,14,15,22,23,30,31,38,39,46,47,54,55,62,63,70,71,78,79},
    ytick={6,7,14,15,22,23,30,31,38,39,46,47,54,55,62,63,70,71,78,79},
    xticklabel= \empty,
    yticklabel=\empty,
    y dir=reverse,
    y grid style={darkgray176},
    xmin=-0.5, xmax=83.5,
    ymin=-0.5, ymax=83.5,
    axis equal image,
    xtick style={color=black, line width=.5pt},
    ytick style={color=black, line width=.5pt}
    ]
    \addplot graphics [xmin=-.5, xmax=83.5, ymin=83.5, ymax=-.5] {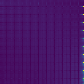};
    \end{axis}
    \end{tikzpicture}
    }%
    \subfloat[Layer 4]{
    \label{subfig:AttExampleLayer4}
    \begin{tikzpicture}[scale=.9]
    \begin{axis}[
    tick align=outside,
    tick pos=both,
    axis line style={draw=none} ,
    xtick={6,7,14,15,22,23,30,31,38,39,46,47,54,55,62,63,70,71,78,79},
    ytick={6,7,14,15,22,23,30,31,38,39,46,47,54,55,62,63,70,71,78,79},
    xticklabel= \empty,
    yticklabel=\empty,
    y dir=reverse,
    y grid style={darkgray176},
    xmin=-0.5, xmax=83.5,
    ymin=-0.5, ymax=83.5,
    axis equal image,
    xtick style={color=black, line width=.5pt},
    ytick style={color=black, line width=.5pt}
    ]
    \addplot graphics [xmin=-.5, xmax=83.5, ymin=83.5, ymax=-.5] {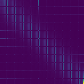};
    \end{axis}
    \end{tikzpicture}
    }%
    \subfloat[Layer 6]{
    \label{subfig:AttExampleLayer6}
    \begin{tikzpicture}[scale=.9]
    \begin{axis}[
    tick align=outside,
    tick pos=both,
    axis line style={draw=none} ,
    xtick={6,7,14,15,22,23,30,31,38,39,46,47,54,55,62,63,70,71,78,79},
    ytick={6,7,14,15,22,23,30,31,38,39,46,47,54,55,62,63,70,71,78,79},
    xticklabel= \empty,
    yticklabel=\empty,
    y dir=reverse,
    y grid style={darkgray176},
    xmin=-0.5, xmax=83.5,
    ymin=-0.5, ymax=83.5,
    axis equal image,
    xtick style={color=black, line width=.5pt},
    ytick style={color=black, line width=.5pt}
    ]
    \addplot graphics [xmin=-.5, xmax=83.5, ymin=83.5, ymax=-.5] {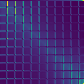};
    \end{axis}
    \end{tikzpicture}
    }
    \caption{\textit{BCJRFormer}'s attention on marker positions increases with layer depth. Shown are the average attention head scores (after applying the $\text{Softmax}$ function in Equation~\eqref{eq:attention}), where brighter colors indicate higher attention. Tick marks align with the positions of markers $\marker = 23$ inserted into a quaternary $(64,32)$ \ac{ldpc} outer code. Results are averaged over 409{,}600 transmissions through an \ac{ids} channel with $\pins = \pdel = 0.01$ and $\psub = 0.012$.}
    \label{fig:attentionExample}
\end{figure}
\section{Attention Visualization}

Figure~\ref{fig:attentionExample} shows attention heatmaps at different depths of the \textit{BCJRFormer} model, which was trained on the quaternary code described in Section~\ref{sec:ExperimentBCJRDecoderSingleMarkerQ4}. We observe that in higher layers, attention is distributed globally. However, the marker positions of Layer 1 (Subfigure~\ref{subfig:AttExampleLayer0}) show very strong attention to the final sequence token. In the \ac{bcjr} algorithm, received sequence length $\nrec$ relative to the codeword length $\nin$ is used to initialize the backward recursion. We conjecture that the model similarly uses the final token to extract early information about the severity of synchronization loss throughout the sequence.

We further note that the final symbol of the sequence is considerably easier to decode than other (non-marker) symbols (see, for example, Figure~\ref{fig:AppQuartErrDistr}), which may further encourage attention. In deeper layers (Subfigures~\ref{subfig:AttExampleLayer4} and~\ref{subfig:AttExampleLayer6}), we observe very local attention that is directed toward nearby marker tokens. This is expected, since the markers provide the main source of prior information that can be used to synchronize the sequence.

\label{sec:appMultipleSequenceAggregation}
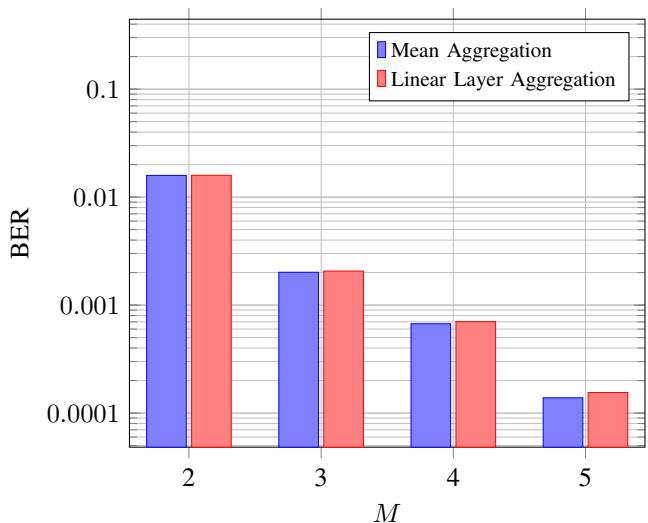
\begin{figure}[h]
    \centering
    \begin{tikzpicture}
        \begin{axis}[
          ybar,   
          ymode=log,
          log origin=infty,
          ymax=0.155,
          grid=both,
          symbolic x coords={2, 3, 4, 5},
          xtick=data,
          grid style={line width=.1pt},
          enlargelimits=0.15,
          legend pos=north east,
          legend cell align={left},
          xticklabel style = log ticks with fixed point,
          xlabel={$M$},
          ylabel={$ \text{BER} $},
          bar width=15pt,              
            legend image code/.code={
                    \draw [#1] (-1pt,-2pt) rectangle (2pt,4pt); },
            ]
          
         
        ]
        
        \addplot [ybar, line width=.2pt, blue, fill=blue!50] coordinates {
            (2, 0.015893)
            (3, 0.002015)
            (4, 0.0006719)
            (5, 0.00013855) 
         };
        \addlegendentry{Mean Aggregation}
        
        \addplot [ybar, line width=.2pt, red, fill=red!50] coordinates {
            (2, 0.015927)
            (3, 0.002067)
            (4, 0.00070211)
            (5, 0.00015511) 
         };
         \addlegendentry{Linear Layer Aggregation}

        \end{axis}
    \end{tikzpicture}
    \caption{Mean aggregation achieves lower error rates than a single linear layer for output dimensionality reduction in joint decoding with \textit{BCJRFormer}. Models are trained on transmissions with $\pins = \pdel = 0.01$ and $\psub = 0.012$ using an outer $(96, 48)$ \ac{ldpc} code with inner markers $\marker =  001$ inserted every $N_m = 6$ bits.}
    \label{fig:ablation_msa_aggregation}
\end{figure}
\section{Ablation: Multiple Sequence Aggregation}

When more than one sequence is transmitted (i.e., $M>1$), the transformer block outputs a tensor of dimension $M\nin$ instead of $\nin$. To reduce the dimension to the inner code size $\nin$, we propose mean aggregation (see Equation~\eqref{eq:mean_dim_reduction}). An alternative is to use a single linear layer. Because a linear layer introduces learnable parameters, it may lead to slight performance improvements. We compare the \ac{ber} of mean aggregation and linear layer aggregation for $M \in \{2, 3, 4, 5\}$. We use the shorter code construction described in Section~\ref{sec:ExperimentBCJRDecoderSingle} and set $\pins = \pdel = 0.01$ and $\psub = 0.012$. Figure~\ref{fig:ablation_msa_aggregation} shows that mean aggregation yields lower error rates for all values of $M$. The performance difference becomes more pronounced as the number of sequences increases. We believe that mean aggregation naturally preserves the significance of each input symbol, whereas a linear layer must learn that every token is meaningful.

\begin{figure}[h]
    \centering
    \begin{tikzpicture}
        \begin{axis}[
          ybar,   
          ymode=log,
          log origin=infty,
          ymax=0.155,
          grid=both,
          symbolic x coords={2, 3, 4, 5},
          xtick=data,
          grid style={line width=.1pt},
          enlargelimits=0.15,
          legend pos=north east,
          legend cell align={left},
          xticklabel style = log ticks with fixed point,
          xlabel={$M$},
          ylabel={$ \text{BER} $},
          bar width=15pt,              
            legend image code/.code={
                    \draw [#1] (-1pt,-2pt) rectangle (2pt,4pt); },
            ]
          
         
        ]
        
        \addplot [ybar, line width=.2pt, blue, fill=blue!50] coordinates {
            (2, 0.015893)
            (3, 0.002015)
            (4, 0.0006719)
            (5, 0.00013855) 
         };
        \addlegendentry{With Sequence Embedding}
        
        \addplot [ybar, line width=.2pt, red, fill=red!50] coordinates {
            (2, 0.016004)
            (3, 0.002143)
            (4, 0.00069712)
            (5, 0.00015645) 
         };
         \addlegendentry{No Sequence Embedding}
        \end{axis}
    \end{tikzpicture}%
    \caption{Adding a sequence embedding improves error rates for jointly decoding multiple noisy copies of a codeword transmitted through a channel with $\pins = \pdel = 0.01$ and $\psub = 0.012$. Shown here for an outer $(96, 48)$ \ac{ldpc} code with markers $\marker = 001$ inserted every $N_m = 6$ bits.}
    \label{fig:appAblationSequenceEmbedding}
\end{figure}
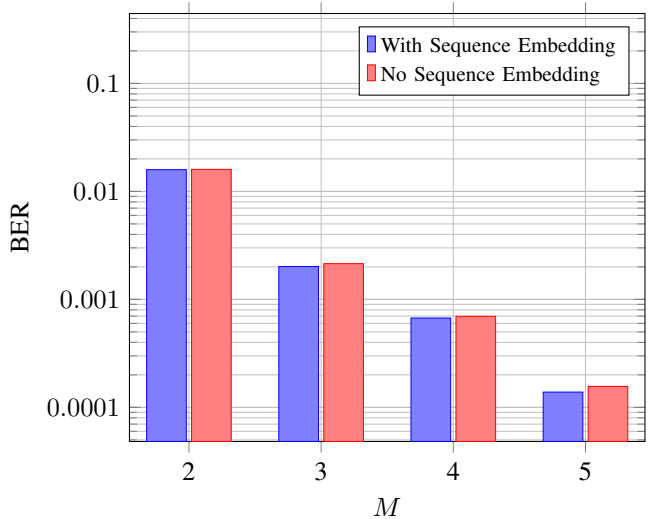
\section{Ablation: Sequence Embedding}
\label{sec:appSequenceEmbedding}
As introduced in Section~\ref{sec:BCJRFormer}, we add a sequence embedding in addition to a positional embedding that indicates the sequence to which each symbol belongs. In Figure~\ref{fig:appAblationSequenceEmbedding}, we compare two models: one without sequence embedding and one with sequence embedding. We use the same code construction as in Section~\ref{sec:ExperimentBCJRDecoderSingle}, namely a $(96, 48)$ \ac{ldpc} code concatenated with markers $\marker=001$ inserted every $\markerFreq = 6$ bits. We fix the channel configuration to $\pins = \pdel = 0.01$ and $\psub = 0.012$ and compare models across $M \in \{2, 3, 4, 5\}$ transmissions. Sequence embedding decreases error rates even for $M=2$, and the figure suggests that the improvement becomes more pronounced as the number of sequences $M$ increases.

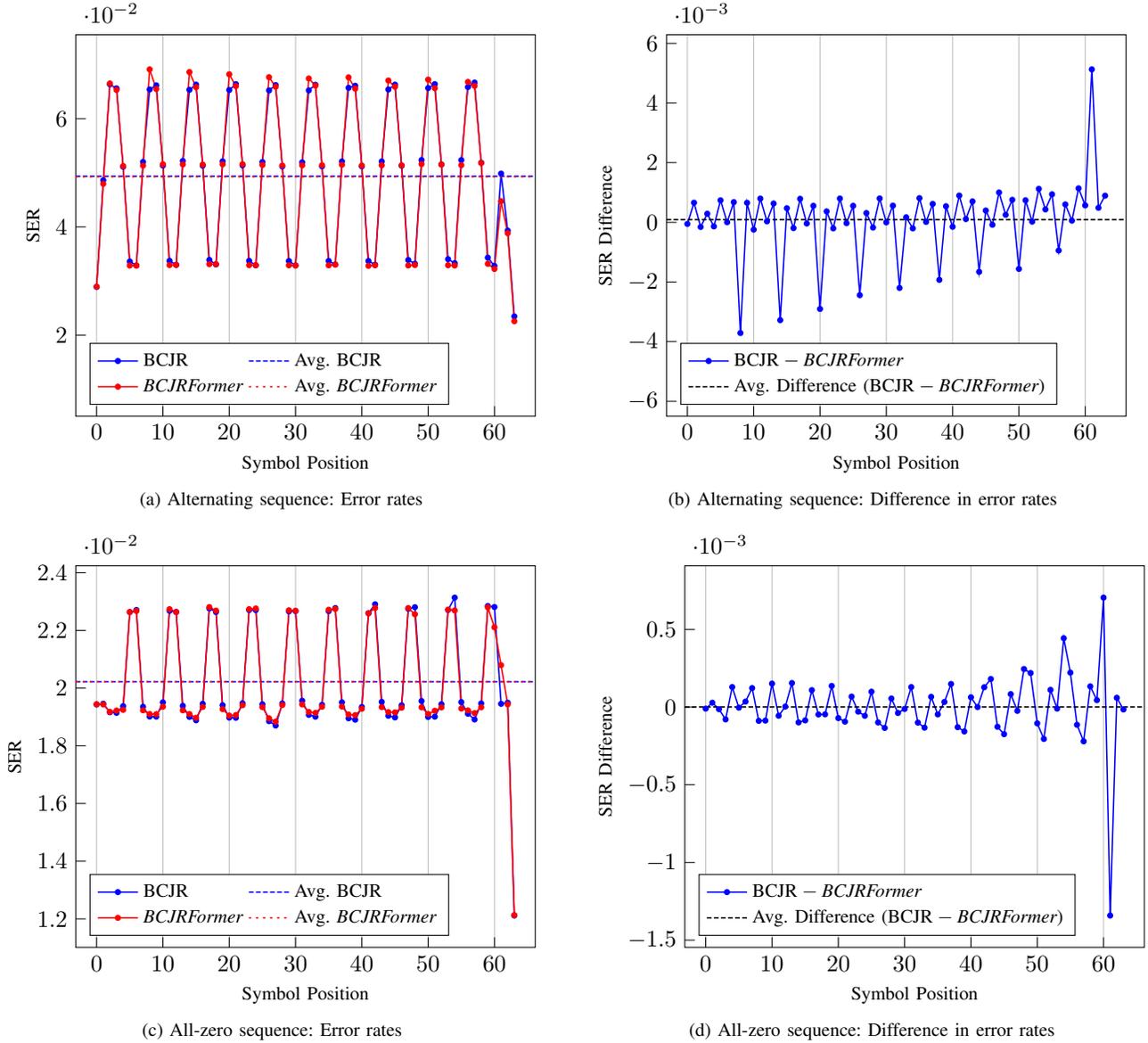
\begin{figure}[ht]
\centering%
\subfloat[Alternating sequence: Error rates]{%
\begin{tikzpicture}
\begin{axis}[
legend cell align={left},
tick align=inside,
tick pos=left,
xlabel={\footnotesize Symbol Position},
xmajorgrids,
xminorgrids,
xmin=-3.15, xmax=66.15,
ymin=0.005,
xtick style={color=black},
ylabel near ticks,
ylabel={\footnotesize SER},
ytick style={color=black},
legend pos=south west,
legend columns=2,
]
\addplot [semithick, blue, mark=*, mark size=1, mark options={solid}]
table {tables/01CodewordQuaternary64ErrBCJR.txt};
\addlegendentry{ BCJR}
\addplot [semithick, blue, dash pattern=on 2pt off 1pt]
table {tables/01CodewordQuaternary64ErrBCJRAvg.txt};
\addlegendentry{ Avg. BCJR}
\addplot [semithick, red, mark=*, mark size=1, mark options={solid}]
table {tables/01CodewordQuaternary64ErrBCJRFormer.txt};
\addlegendentry{ \textit{BCJRFormer}}
\addplot [semithick, red, dash pattern=on 1pt off 2pt]
table {tables/01CodewordQuaternary64ErrBCJRFormerAvg.txt};
\addlegendentry{ Avg. \textit{BCJRFormer}}
\end{axis}
\label{subfig:00errBCJRvsBCJRFormer}
\end{tikzpicture}
}%
\qquad
\subfloat[Alternating sequence: Difference in error rates]{%
\begin{tikzpicture}
\begin{axis}[
legend cell align={left},
tick align=inside,
tick pos=left,
xlabel={ \footnotesize Symbol Position},
xmajorgrids,
xminorgrids,
xmin=-3.15, xmax=66.15,
ymin=-0.0065,
xtick style={color=black},
ylabel near ticks,
ylabel={\footnotesize SER Difference},
ytick style={color=black},
legend pos=south west,
]
\addplot [semithick, blue, mark=*, mark options={solid}, mark size=1]
table {tables/01CodewordQuaternary64ErrDiff.txt};
\addlegendentry{ $\text{BCJR} - \textit{BCJRFormer}$}
\addplot [semithick, black, dash pattern=on 2pt off 1pt]
table {tables/01CodewordQuaternary64ErrDiffAvg.txt};
\addlegendentry{ Avg. Difference ($\text{BCJR} - \textit{BCJRFormer}$)}

\end{axis}
\label{subfig:00errDiffBCJRminusBCJRFormer}
\end{tikzpicture}
}%
\\
\subfloat[All-zero sequence: Error rates]{%
\begin{tikzpicture}
\begin{axis}[
legend cell align={left},
tick align=inside,
tick pos=left,
xlabel={\footnotesize Symbol Position},
xmajorgrids,
xminorgrids,
xmin=-3.15, xmax=66.15,
xtick style={color=black},
ylabel near ticks,
ylabel={\footnotesize SER},
ytick style={color=black},
legend columns=2,
legend pos= south west,
]
\addplot [semithick, blue, mark=*, mark size=1, mark options={solid}]
table {tables/00CodewordQuaternary64ErrBCJR.txt};
\addlegendentry{BCJR}
\addplot [semithick, blue, dash pattern=on 2pt off 1pt]
table {tables/00CodewordQuaternary64ErrBCJRAvg.txt};
\addlegendentry{Avg. BCJR}
\addplot [semithick, red, mark=*, mark size=1, mark options={solid}]
table {tables/00CodewordQuaternary64ErrBCJRFormer.txt};
\addlegendentry{\textit{BCJRFormer}}
\addplot [semithick, red, dash pattern=on 1pt off 2pt]
table {tables/00CodewordQuaternary64ErrBCJRFormerAvg.txt};
\addlegendentry{Avg. \textit{BCJRFormer}}
\end{axis}
\label{subfig:01errBCJRvsBCJRFormer}
\end{tikzpicture}
}%
\qquad
\subfloat[All-zero sequence: Difference in error rates]{%
\begin{tikzpicture}
\begin{axis}[
legend cell align={left},
tick align=inside,
tick pos=left,
xlabel={\footnotesize Symbol Position},
xmajorgrids,
xmin=-3.15, xmax=66.15,
xminorgrids,
xtick style={color=black},
ylabel near ticks,
ylabel={\footnotesize SER Difference},
ytick style={color=black},
legend pos=south west,
]
\addplot [semithick, blue, mark=*, mark size=1, mark options={solid}]
table {tables/00CodewordQuaternary64ErrDiff.txt};
\addlegendentry{$\text{BCJR} - \textit{BCJRFormer}$}
\addplot [semithick, black, dash pattern=on 2pt off 1pt]
table {tables/00CodewordQuaternary64ErrDiffAvg.txt};
\addlegendentry{Avg. Difference ($\text{BCJR} - \textit{BCJRFormer}$)}
\end{axis}
\label{subfig:01errDiffBCJRminusBCJRFormer}
\end{tikzpicture}
}%
\caption{
Error distributions for the alternating and all-zero sequences\,---\,shown in Subfigures \protect\subref{subfig:00errBCJRvsBCJRFormer} and \protect\subref{subfig:01errBCJRvsBCJRFormer}, respectively\,---\,differ at specific positions between \textit{BCJRFormer} and \ac{bcjr} at specific positions, despite similar overall error rates. Subfigures \protect\subref{subfig:00errDiffBCJRminusBCJRFormer} and \protect\subref{subfig:01errDiffBCJRminusBCJRFormer} show the error rate differences, where values $>0$ indicate that \textit{BCJRFormer} outperforms \ac{bcjr} and values $<0$ indicate the reverse. The alternating and all-zero sequence of length $64$, concatenated with markers $\marker = 32$ inserted every $\markerFreq=6$ symbols, were transmitted through a channel with $\psub = 0.012$ and $\pdel = \pins = 0.01$. 
}
\label{fig:AppQuartErrDistr}
\end{figure}
\section{Output Distribution: Comparison of BCJR and BCJRFormer}
\label{sec:appBCJRFormerQuartDistribution}

For transmissions over a deletion channel, the size of the deletion error balls (i.e., the number of sequences that can arise by deleting a fixed number of symbols) increases with the number of runs in the transmitted sequence. A run is defined as a block of consecutive symbols in a sequence (for example, the sequence $002111$ has three runs). A maximum likelihood decoder can distinguish between two received sequences if their corresponding error balls do not intersect. Intuitively, transmitting an alternating codeword (e.g. $010101\ldots$) is more difficult to synchronize than transmitting the all-zero codeword because the alternating sequence has $\nin$ runs, while the all-zero sequence has only one run~\cite{sloaneSingledeletioncorrectingCodes2002a, mitzenmacherSurveyResultsDeletion2009a}. 

We explore the symbol-wise error rate distribution of \textit{BCJRFormer} by comparing it to the \ac{bcjr} algorithm for transmissions of these two codewords in the quaternary domain.
We concatenate each sequence of length $64$ with markers $\marker = 32$ inserted every $\markerFreq = 6$ symbols, and transmit them via an \ac{ids} channel with $\pins = \pdel = 0.01$ and $\psub = 0.012$. We reuse the corresponding model from Section~\ref{sec:ExperimentBCJRDecoderSingleMarkerQ4} and consider the same 40{,}960{,}000 transmissions of each sequence for both decoders. Figure~\ref{fig:AppQuartErrDistr} shows (left column) the \acp{ser} for \textit{BCJRFormer} and the \ac{bcjr} algorithm, as well as the difference between the error rate of \ac{bcjr} and that of \textit{BCJRFormer} (right column). As expected, the average error rates of the alternating sequence (first row) are much higher than those of the all-zero sequence (second row). The error rate of symbols is strongly correlated with the distance from the nearest inserted marker.

\section{Ablation: Failure to Decode in Low Probability Domain for Vectorized Inputs}
\label{sec:appInputAggregation}
\begin{figure}[h]
\centering
\begin{tikzpicture}
    \begin{axis}[
        scaled ticks=true,
        height=7cm,
        ymin=0.001,
        scaled ticks=false, 
        xticklabel style = {/pgf/number format/fixed, /pgf/number format/precision=3},
        yticklabel style = log ticks with fixed point,
        xtick distance=0.005,
        ymode=log,
        grid = both,
        grid style = {line width=.1pt},
        legend pos=south east,
        tick align=inside,
        legend cell align={left},
        xlabel = {$\pins=\pdel$},
        ylabel = {BER},
        ]

    \addplot [mark=square, mark options={solid}, mark size=1.5pt,color=blue] 
    coordinates {
        (0.001, 0.33853)
        (0.002, 0.17931)
        (0.003, 0.023747)
        (0.004, 0.026499)
        (0.005, 0.023745)
        (0.006, 0.026431)
        (0.007, 0.029081)
        (0.008, 0.026541)
        (0.009, 0.029162)        
        (0.01, 0.031862)
        (0.015, 0.039683)
     };
    \addlegendentry{Aggregated Input}

    \addplot [mark=square, mark options={solid}, mark size=1.5pt, color=red] coordinates { 
        (0.001, 0.0027213)
        (0.002, 0.0054509)
        (0.003, 0.0080844)
        (0.004, 0.010809)
        (0.005, 0.013461) 
        (0.006, 0.016145)
        (0.007, 0.018826)
        (0.008, 0.021366)
        (0.009, 0.023998)     
        (0.01, 0.026641) 
        (0.015, 0.039711)
    };
    \addlegendentry{Standard Input}

    \end{axis}
\end{tikzpicture}%
\caption{Models using aggregated inputs fail to synchronize sequences when error probabilities are very low. The figure shows results for transmissions at $\psub = 0.012$ using a \((96, 48)\) \ac{ldpc} outer code with markers $\marker = 001$ inserted at fixed intervals \(\markerFreq = 6\).}
\label{fig:appInputAggregation}
\end{figure}
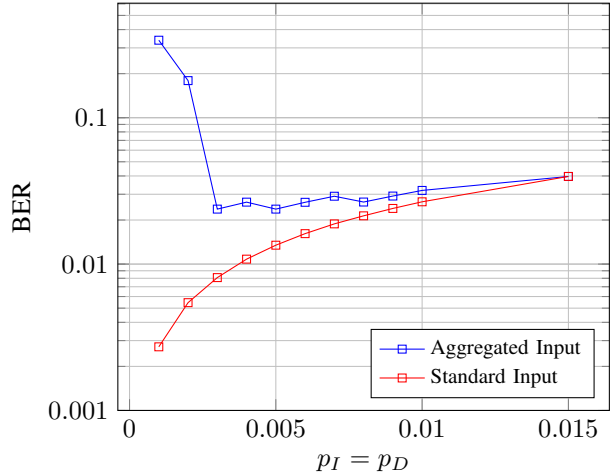

In Section~\ref{sec:methBCJRFormerSingleTransmission}, we introduced the tensor input for \textit{BCJRFormer}. More specifically, for $1 \leq i \leq \nin$, we presented the construction of matrices $\bcjrformerInput_i \in \mathbb{R}^{\delta \times q}$ in Equation~\eqref{eq:BCJRFormerInput}. For binary marker codes, we consider an alternative construction in which the input is summed. In this construction, we define $\bcjrformerInput_i^{v} \in \mathbb{R}^{\delta}$ as 
 \begin{equation}
     \label{eq:AppBCJRFormerInputAgg}
    \bcjrformerInputIx^{v}_{i,j} = \sum_{\Yval \in \sbin} P(\yinIx_i = \Yval)F(\Yval, \recIx_{i + d_j}).
 \end{equation} 
For non-marker positions $i$, we have $\bcjrformerInputIx^{v}_{i,j} = 0.5$. For marker symbols with the value $\Yval_m$, we have $\bcjrformerInputIx^{v}_{i,j} = F(\Yval_m, \recIx_{i + d_j})$. If every non-marker symbol is covered by the window associated with a marker symbol (i.e., $\delta \geq \markerFreq$), it is unclear whether the aggregated input representation performs worse than the proposed one. For $\pins = \pdel \geq 0.015$, the aggregated input performs equally well to the proposed one. Figure~\ref{fig:appInputAggregation} shows the error rates for $\pins = \pdel \leq 0.015$ at $\psub = 0.012$ for models trained with the proposed input and with the aggregated input. We reused the model and shorter concatenated code construction from Section~\ref{sec:ExperimentBCJRDecoderSingle}.

Models trained on the aggregated input generally perform worse and do not improve for deletion and insertion probabilities below $0.01$. Furthermore, their performance deteriorates considerably for probabilities $\pdel = \pins < 0.003$. We conjecture that the vectorized input leads to a much sparser representation of synchronization errors. When combined with a very low probability of synchronization errors, this results in the model having insufficient error representation to learn how to synchronize general sequences effectively. Our proposed input does not have this issue, because it also captures synchronization errors within tokens that are not at marker positions.

\section{Training Convergence of ConvBCJRFormer With and Without Masking}
\label{sec:appTrainConvergenceOfConvBCJRFormer}

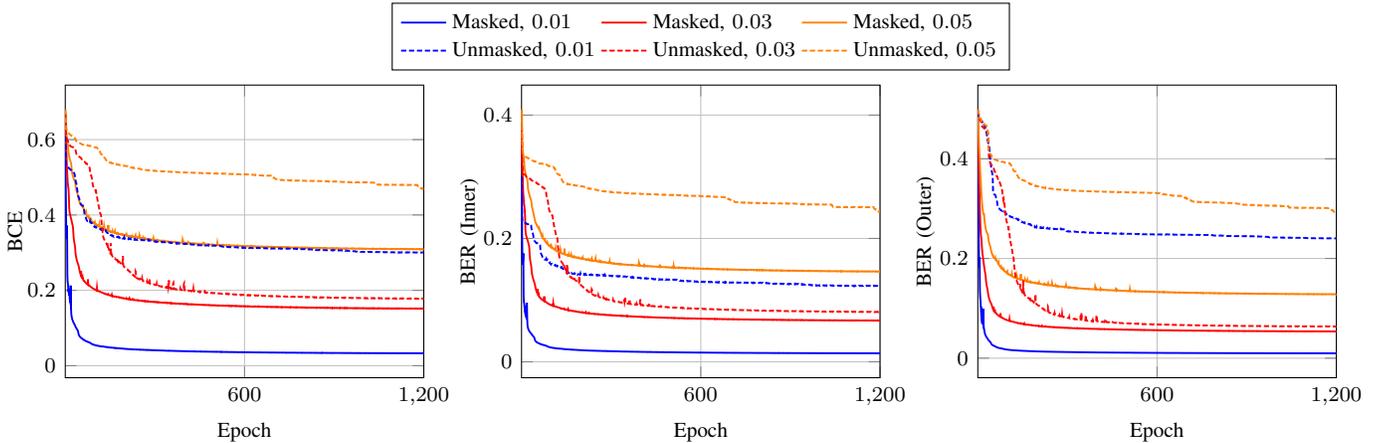
\begin{figure}[ht!]
    \centering
    \begin{tikzpicture}
    \tikzstyle{every node}=[font=\footnotesize]
    \begin{groupplot}[
        group style={group size=3 by 1, horizontal sep=1.3cm}, 
        width=.35\textwidth,
        xtick={600, 1200}, 
        xmin=0, 
        xmax=1200,
        xticklabel style = log ticks with fixed point,
        grid = both, 
        ylabel style={
            yshift=-5pt,
        },
        grid style = {line width=.1pt}, 
        tick align=inside, 
        xlabel = {Epoch},
        every axis plot post/.append style={mark=\empty, line width=.7pt},
    ]
    \nextgroupplot[
        ylabel = {BCE},
        legend to name={CommonLegend},
        legend style={legend columns=3},
        legend cell align={left},
    ]
    \addplot [color=blue] table [x=Epoch, y=TrainLoss, col sep=comma] {tables/convformer_train_loss_001_ps_0012.csv};
    \addlegendentry{Masked, $0.01$}
    
    \addplot [color=red] table [x=Epoch, y=TrainLoss, col sep=comma] {tables/convformer_train_loss_003_ps_0012.csv};
    \addlegendentry{Masked, $0.03$}
    
    \addplot [color=orange] table [x=Epoch, y=TrainLoss, col sep=comma] {tables/convformer_train_loss_005_ps_0012.csv};
    \addlegendentry{Masked, $0.05$}

    \addplot [dashed, color=blue, dash pattern=on 2pt off 1pt] table [x=Epoch, y=TrainLoss, col sep=comma] {tables/convformer_train_loss_001_ps_0012_unmasked.csv};
    \addlegendentry{Unmasked, $0.01$}
    
    \addplot [dashed, color=red, dash pattern=on 2pt off 1pt] table [x=Epoch, y=TrainLoss, col sep=comma] {tables/convformer_train_loss_003_ps_0012_unmasked.csv};
    \addlegendentry{Unmasked, $0.03$}
    
    \addplot [dashed, color=orange, dash pattern=on 2pt off 1pt] table [x=Epoch, y=TrainLoss, col sep=comma] {tables/convformer_train_loss_005_ps_0012_unmasked.csv};
    \addlegendentry{Unmasked, $0.05$}

    \nextgroupplot[
        ylabel = {BER (Inner)},
    ]
    \addplot [color=blue] table [x=Epoch, y=BERInner, col sep=comma] {tables/convformer_train_ber_inner_001_ps_0012.csv};
    
    \addplot [color=red] table [x=Epoch, y=BERInner, col sep=comma] {tables/convformer_train_ber_inner_003_ps_0012.csv};
    
    \addplot [color=orange] table [x=Epoch, y=BERInner, col sep=comma] {tables/convformer_train_ber_inner_005_ps_0012.csv};

    \addplot [dashed, color=blue, dash pattern=on 2pt off 1pt] table [x=Epoch, y=BERInner, col sep=comma] {tables/convformer_train_ber_inner_001_ps_0012_unmasked.csv};
    
    \addplot [dashed, color=red, dash pattern=on 2pt off 1pt] table [x=Epoch, y=BERInner, col sep=comma] {tables/convformer_train_ber_inner_003_ps_0012_unmasked.csv};
    
    \addplot [dashed, color=orange, dash pattern=on 2pt off 1pt] table [x=Epoch, y=BERInner, col sep=comma] {tables/convformer_train_ber_inner_005_ps_0012_unmasked.csv};
    
    \nextgroupplot[
        ylabel = {BER (Outer)},
    ]
    \addplot [color=blue] table [x=Epoch, y=BEROuter, col sep=comma] {tables/convformer_train_ber_outer_001_ps_0012.csv};
    
    \addplot [color=red] table [x=Epoch, y=BEROuter, col sep=comma] {tables/convformer_train_ber_outer_003_ps_0012.csv};
    
    \addplot [color=orange] table [x=Epoch, y=BEROuter, col sep=comma] {tables/convformer_train_ber_outer_005_ps_0012.csv};

    \addplot [dashed, color=blue, dash pattern=on 2pt off 1pt] table [x=Epoch, y=BEROuter, col sep=comma] {tables/convformer_train_ber_outer_001_ps_0012_unmasked.csv};
    
    \addplot [dashed, color=red, dash pattern=on 2pt off 1pt] table [x=Epoch, y=BEROuter, col sep=comma] {tables/convformer_train_ber_outer_003_ps_0012_unmasked.csv};
    
    \addplot [dashed, color=orange, dash pattern=on 2pt off 1pt] table [x=Epoch, y=BEROuter, col sep=comma] {tables/convformer_train_ber_outer_005_ps_0012_unmasked.csv};
    
    \end{groupplot}
    
    \path (group c2r1.north west) -- node[above, yshift=2pt]{\begin{NoHyper}\ref{CommonLegend}\end{NoHyper}} (group c2r1.north east);
    
    \end{tikzpicture}
    \caption{Unmasked models fail to capture the convolutional structure. We compare metrics for insertion/deletion probabilities $\pins = \pdel \in \{0.01, 0.03, 0.05\}$ at a fixed substitution probability of $\psub =0.012$. The left graph shows the overall \ac{bce} error, the middle graph the \ac{ber} between the inner codeword $\yin$ and its prediction $\predyin$, and the right graph the \ac{ber} between the outer codeword $\yout$ and its prediction $\predyout$. One epoch corresponds to $400$ iterations.}
    \label{fig:appAblationMasking}
\end{figure}

We evaluate the impact of our proposed masking of \textit{ConvBCJRFormer}'s cross-attention mechanism using the generator matrix $G$. In Figure~\ref{fig:appAblationMasking}, we compare the training convergence of models from Experiment~\ref{sec:ExpConvBCJRFormer} with equivalent models trained without masking. We observe that using the generator matrix $G$ to exclude unrelated symbols and states from the attention mechanism yields much better final error rates. Furthermore, we observe that for certain probability ranges\,---\,such as when $\pins = \pdel = 0.01$\,---\,the training of unmasked models converges to a suboptimal state, presumably a local minimum. This premature convergence results in higher error rates than those observed for models trained on sequences with higher deletion/insertion error probabilities.

\section{LDPC quaternary Protograph Construction}
\label{sec:appQuartProtographConstruction}
\begin{figure}[h]
    \centering
    \begin{tikzpicture}[auto]
      
      \node[vnode] (v1) at (0,0) {$v_1$};
      \node[vnode, right = 10pt of v1.east] (v2) {$v_2$};
      \node[vnode, right = 10pt of v2.east] (v3) {$v_3$};
      \node[vnode, right = 10pt of v3.east] (v4) {$v_4$};
      
      \node[cnode, above right = 20pt and 10pt of v1.north east] (c1) {$c_1$};
      \node[cnode, above left = 20pt and 10pt of v4.north west] (c2) {$c_2$};
    
      \path [rounded corners=.96cm, name path=A] (c1.south)|-(c1.east);
      \path [name path=B] (c1.east -| c1.south) -- (c1.south -| c1.east);
      \coordinate[name intersections={of=A and B, by=c1SE}];

    \path (v1) edge (c1.-115);
    \path (v2.111) edge (c1.-101);
    \path (v2.97) edge (c1.-87);
    \path (v3) edge (c1.-73);
    \path (v4) edge (c1.-59);
    
    \path (v1) edge (c2.-115);
    \path (v2) edge (c2.-101);
    \path (v3.83) edge (c2.-87);
    \path (v3.69) edge (c2.-73);
    \path (v4) edge (c2.-759);
      
    \end{tikzpicture}
    \caption{Protograph $\bm{P}$ used to construct the parity-check matrix $\pc$ in Section~\ref{sec:ExperimentBCJRDecoderSingleMarkerQ4}}
    \label{fig:ldpc-protograph}
\end{figure}
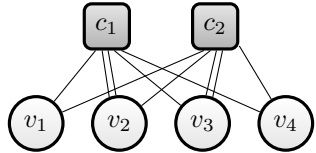

For reproducibility, we present the protograph used in the experiments in Section~\ref{sec:ExperimentBCJRDecoderSingleMarkerQ4}. A protograph is a small Tanner graph, given by a matrix $\bm{P}$ with $m_p$ rows corresponding to check nodes and $n_p$ columns corresponding to variable nodes. Each entry $p_{i,j}$ gives the number of edges between check node $i$ and variable node $j$. For our experiment, we use the protograph proposed in the paper~\cite{maaroufConcatenatedCodesMultiple2023a}, given by 
$$
    \bm{P} = \begin{bmatrix}
        1 & 2 & 1 & 1 \\
        1 & 1 & 2 & 1 \\
    \end{bmatrix}.
$$
Figure~\ref{fig:ldpc-protograph} shows the corresponding graph. A full parity-check matrix is derived by lifting the protograph. First, the protograph is replicated a specified number of times; then, the edges are permuted so that both the degree and interconnectivity between check nodes and variable nodes are preserved~\cite{butlerBoundsMinimumDistance2013}. For non-binary codes, the weights of the edges in the resulting parity-check matrix can be chosen randomly. In our experiments, we randomly initialized the weights once and then fixed them for all experiments.

\end{document}